\documentclass[]{aa} % for the letters 
\usepackage{graphicx}
\usepackage{longtable}
\usepackage{lscape}
\usepackage{gensymb}
\usepackage{txfonts}
\begin{document}

   \title{Size and albedo distributions of asteroids in cometary orbits using WISE data}

%   \subtitle{I. Overviewing the $\kappa$-mechanism}

   \author{J. Licandro\inst{1,2}
             \and
          V. Al\'i-Lagoa\inst{1,2}
          \and
          G. Tancredi\inst{3}
          \and
          Y. Fern\'andez\inst{4}
          }

   \institute{Instituto de Astrof\'isica de Canarias (IAC),
             C/V\'ia L\'actea s/n, 38205 La Laguna, Tenerife, Spain.\\
              \email{jlicandr@iac.es}
         \and
             Departamento de Astrof\'{\i}sica, Universidad de La Laguna (ULL), 
             E-38205 La Laguna, Tenerife, Spain.\\
         \and
             Departamento de Astronom\'{\i}a, Facultad de Ciencias, Igu\'a 4225, 11400 Montevideo, Uruguay.\\
         \and
              Physics Department, University of Central Florida, P. O. Box 162385, Orlando, FL 32816.2385, USA
     }

   \date{Received XXX xx, XXX; accepted XXX xx, XXX}

% \abstract{}{}{}{}{} 
% 5 {} token are mandatory
 
  \abstract
  {Determining whether asteroids in cometary orbits (ACOs)  are dormant or extinct comets is relevant for understanding the end-states of comets and the sizes of the comet population.}
 %  {Main belt comets (MBCs) are object in orbits typical of main belt asteroids that have been observed with a comet-like coma and/or tail. To understand the mechanism of activation and their origin provide information of cosmogonical interest.  }
  {We intend to study the value distributions of effective diameter ($D$), beaming parameter ($\eta$), and visible geometric albedo ($p_V$) of ACO populations, which can be derived from NASA's Wide-field Infrared Explorer (WISE) observations, and we aim to compare these with the same, independently determined properties of the comets.}
   {The near-Earth asteroid thermal model (NEATM) is used  with WISE data and the absolute magnitude ($H$) of the ACOs to compute the $D$, $p_V$ and $\eta$.}
   {
     We obtained $D$ and $p_V$ for 49 ACOs in Jupiter family cometary orbits (JF-ACOs) and 16 ACOs in Halley-type cometary orbits (Damocloids). We also obtained  the infrared beaming parameter $\eta$ for 45 of them. All but three JF-ACOs (95\% of the sample) present a low albedo compatible with a cometary origin. The $p_V$ and $\eta$ distributions of both ACO populations are very similar. For the entire sample of ACOs, the mean geometric albedo  is $\bar{p_V} = 0.05 \pm 0.02$, ($\bar{p_V} = 0.05 \pm 0.01$ and $\bar{p_V} =0.05 \pm 0.02$ for JF-ACOs and for Damocloids, respectively) compatible with a narrow albedo distribution similar to that of the Jupiter family comets (JFCs), with a $\bar{p_V} \sim 0.04$. %around  $\bar{p_V} \sim 0.045$ which is 
     The mean beaming parameter is $\bar{\eta} =1.0 \pm 0.2$. We find no correlations between $D$, $p_V$, or $\eta$.  %We search for correlations between  $D$, $p_V$, $\eta$ and the phase angle of the observations ($\theta$). Only a linear relation between $\eta$ and $\theta$ ($\eta=0.85 + 0.008 \ \theta$) is found in our data. 
     We also compare  the cumulative size distribution (CSD) of ACOs, Centaurs, and JFCs. Although the Centaur sample contains larger objects, the linear parts in their log-log plot of the CSDs presents a similar cumulative exponent ($\beta = 1.85 \pm 0.30$ and $1.76 \pm 0.35$, respectively). The CSD for Damocloids presents a much shallower exponent $\beta = 0.89 \pm 0.17$.}
   {The $p_V$- and $\eta$-value distributions of ACOs are very similar to those of JF comet (JFCs) nuclei. The ACOs in Tancredi's list are the best possible candidates to be dormant/inactive comets. The CSD for JF-ACOs is shallower and shifted towards larger diameters with respect to the CSD of active JFCs, which suggests that the mantling process has a size dependency whereby large comets tend to reach an inactive stage faster than  small comets. Finally, the population of JF-ACOs is comparable in number with the population of JFCs, although there are more tens-km JF-ACOs than JFCs.}
   

   \authorrunning{Licandro, J. et \emph{al.}}

   \titlerunning{Size and albedo distributions of ACOs using WISE data}
   
   \maketitle
%
%________________________________________________________________

%%%%%%%%%%%%%%%%%%%%%%%%%%%
%%%%%%%%%%%%%%%%%%%%%%%%%%%
\section{Introduction}
%%%%%%%%%%%%%%%%%%%%%%%%%%%
%%%%%%%%%%%%%%%%%%%%%%%%%%%

The classification criterion between asteroids and comets has evolved in recent decades, but the main phenomenological distinction remains unchanged: comets are active objects in the sense that they show gas and/or dust ejected from the surface at some point in their orbits, whereas asteroids do not. This statement is purely observational and makes no distinction among the different ways in which a body can lose mass.

There are some borderline cases between the asteroid and comet populations, usually called comet-asteroid transition objects, which include: (1) objects in typical cometary orbits that have never shown (or have not recently shown) any kind of activity, the ``asteroids in cometary orbits'' (ACOs); and (2) objects in typical asteroidal orbits that present some evidence of dust ejection, the ``active asteroids'' (AAs), a group that includes the so-called ``main belt comets'' (MBCs).  Some AAs are designated comets because of the observed dust ejection, even though further analysis may later indicate that the dust is not produced by traditional cometary activity (i.e. sublimation or phase transition of volatiles), but from other physical phenomena, such as rotational spin-up, collisions, or thermal cracking. 

To identify the transitional objects, several classification schemes have been used that are usually based on Tisserand's  parameter ($T_J$, Kres\'ak \cite{kresak1979}). $T_J$ is a constant of motion in the restricted three-body problem and is related to an object's encounter velocity with Jupiter. Most comets have orbits with $T_J<3$, while the great majority of asteroids have orbits with $T_J>3$. Only a few known comets, about 20 in the Jupiter family, have $T_J>3$, with 2P/Encke being the most well known, but this is attributed to the long-term effects of non-gravitational forces (Fern\'andez et al. \cite{Fernandez2002}). Among asteroids with $T_J<3$, aside from the ACOs mentioned above, are some members of the Cybele, Hilda, and Jupiter Trojan populations, and other objects close to the mean-motion resonances. 

The problem with the usual $T_J$ criterion is that it includes many objects with stable dynamical evolution incompatible with the chaotic dynamics of comets. Tancredi \cite{Tancredi2014} presented a much more restrictive criterion to identify ACOs, which ensures the selection of objects with a dynamical evolution similar to the population of periodic comets. Tancredi's criterion is based on Tisserand's parameter, the minimum orbital intersection distance (MOID) among the giant planets, and some information regarding the aphelion and perihelion distances, and it does not include objects in mean-motion resonances. 

After applying a filter to the sample of $>$500,000 known asteroids to select only those with precise-enough orbits, Tancredi \cite{Tancredi2014} applied the proposed classification criterion to identify the ACOs. The resulting sample consists of $\sim$331 ACOs, which are further classified in subclasses similar to the cometary classification: 203 objects belong to the Jupiter family group; 72 objects are classified as Centaurs; and 56 objects have Halley-type orbits (also known as ``Damocloids''). These are the best known extinct/dormant comet candidates from a dynamical point of view.

To determine whether these objects truly are extinct/dormant comets or simply asteroids that have escaped from the main belt into cometary-like chaotic orbits, observations of other physical properties are needed. Recent studies of the surface properties of ACOs, including photometric and spectroscopic studies in the visible and near-infrared (Licandro et al.,  \cite{licandro2006}; Alvarez-Candal and Licandro, \cite{AlvarezCandal2006}; DeMeo and Binzel, \cite{DeMeo2008}; Licandro et al., \cite{licandro08}; Alvarez-Candal, \cite{AlvarezCandal2013}) and in the mid-infrared (Fern\'andez et al., \cite{Fernandez05};  Licandro et al. \cite{licandro09}), feature a significant number of asteroids that do not satisfy Tancredi's criterion. For example, Licandro et al. (\cite{licandro08}) studied the visible and near-IR spectral properties of 39 ACOs, but only six appear in Tancredi's list; in fact, based on their spectral properties, Licandro et al. (\cite{licandro08}) warned that there are a significant number of objects in their sample that are likely ``interlopers''. DeMeo and Binzel (\cite{DeMeo2008}) analysed the spectra of 55 ACOs (see their Table 3), and after removing six objects from the original list that were already identified as comets, only six out of the remaining 49 objects appear on Tancredi's list. Fern\'andez et al. (\cite{Fernandez05}) computed the albedos of 26 ACOs using mid-IR observations. Two of them are presently classified as comets, and of the remaining 24 objects, 12 are on Tancredi's list. In a recent paper, Kim et al. (\cite{Yoonyoung2014}) presented a compilation of the diameters and albedos of 123 ACOs obtained from different published databases, using their own selection criteria of ACOs; only 29 of these objects are on Tancredi's list. Thus, the number of ACOs with spectral and thermal properties known is very small and the present knowledge of the physical properties of ACOs should be revised in view of Tancredi's results.

The aim of this paper is to study the physical properties of ACOs, which can be derived by fitting a thermal model to  NASA's Wide-field Infrared Survey Explorer (WISE) observations. In particular, we present here the effective diameters ($D$), beaming parameters ($\eta$), visible geometric albedos ($p_V$), and  infrared-to-visible albedo ratios ($R_{\mathrm{W1}}$) using the method described in Al\'i-Lagoa et al (\cite{Ali2013}). We compare our results to previous mid-infrared surveys of ACOs, such as that of Fern\'andez et al. (\cite{Fernandez05}), mentioned earlier, who used the near-Earth asteroid thermal model  (NEATM; Harris 1998) to compute effective radii. Fern\'andez et al.  defined a comet-like albedo as being below 0.075, and they found that 64\% $\pm$ 5\% of their sample has comet-like albedos. They also found a very strong trend of albedo with $T_J$, with nearly all objects in their sample with $T_J< 2.6$ (and thus a higher probability of cometary origin) having comet-like albedos. Licandro et al. (\cite{licandro08}) also found that the spectra of all objects with $T_J<2.7$ in their sample are D- or P-type, which is typical of cometary nuclei.

We present results on ACOs that are part of two of the three populations in Tancredi (\cite{Tancredi2014}), those in Jupiter family comet orbits (JF-ACOs) and those in Halley-type orbits (the Damocloids), assuming they likely have an origin in two different comet reservoirs, the trans-neptunian belt and the Oort cloud, respectively. The WISE data for the population of Centaurs has already been analysed by Bauer et al. \cite{Bauer2013}; we combine our results with theirs to compare the cumulative size distributions (CSDs) of the three populations.

The paper is organized as follows. In Sect. \ref{sec:data}, we briefly describe the WISE data set, the selection criteria and the thermal modelling used to derive $D$, $\eta,$ and $p_V$. We present our results in Sect. \ref{sec:results}, which includes the $p_V$ and $\eta$ distributions. The cumulative size distributions of the three ACO populations are presented in Sec. \ref{sec:sizes} and compared to that of Jupiter family comets. Finally, the discussion and conclusions are presented in Sect. \ref{sec:discussion}.

%%%%%%%%%%%%%%%%%%%%%%%%%%%
%%%%%%%%%%%%%%%%%%%%%%%%%%%
\section{Data and thermal modelling \label{sec:data}}
%%%%%%%%%%%%%%%%%%%%%%%%%%%
%%%%%%%%%%%%%%%%%%%%%%%%%%%

The Wide-field Infrared Survey Explorer (WISE) is a space telescope launched by NASA in December 2009, which carried out an all-sky survey using four broadband filters with approximate isophotal wavelengths at 3.4, 4.6, 12, and 22 $\mu$m (named W1, W2, W3, and W4, respectively); see Wright et al. \cite{Wright2010} and references therein for a general description of the mission. The NEOWISE project managed the detection and archiving of all the observations of the more than  158\,000 small solar system bodies acquired by WISE (see Mainzer et al. \cite{Mainzer2011a}), and all of these observations have been reported to the IAU Minor Planet Center (MPC).  The WISE All-Sky Single Exposure L1b Source Table, available via the IRSA/IPAC archive\footnote{http://irsa.ipac.caltech.edu/Missions/wise.html}, includes the corresponding magnitudes and uncertainties in the Vega system, as well as quality, contamination, and confusion flags that enable us to reject defective data (Cutri et al. \cite{Cutri2012}). Our rejection criteria, summarized next, are based on a combination of those used by Mainzer et al. (\cite{Mainzer2011b}, \cite{Mainzer2011c}), Masiero et al.  (\cite{Masiero2011}), and Grav et al. (\cite{Grav2012}). 

We downloaded all the data matching the MPC reported detections within a cone search radius of 1\arcsec\, centred on the MPC ephemeris of the ACO, and rejected those with artifact flags other than p, P, and 0 and quality flags other than A, B, and C. We also required the modified Julian date to be within four seconds of the time specified in the MPC. We ensured that the data are not contaminated by inertial sources by removing those points that return a positive match from the WISE Source Catalog within 6\arcsec\ if the flux of the inertial source is larger than 5\% of that of the asteroid. We did not use data saturated to any extent, which translates to rejecting objects with magnitudes $M_{\mathrm{W1}} < 7.8$, $M_{\mathrm{W2}} < 6.4$, $M_{\mathrm{W3}} < 3.6$, or $M_{\mathrm{W4}} < -0.6$ (Cutri et al. \cite{Cutri2012}). All remaining observations in a given band are rejected if they are fewer than 40\% of the data in the band with the maximum number of detections. Finally, we implemented the correction to the W3 and W4 isophotal wavelengths and zero magnitude fluxes to account for the red and blue calibrator discrepancy in W3 and W4 (Wright et al. \cite{Wright2010}). The application of the above criteria results in a sample of 49 JF-ACOs  and 16 Damocloids with WISE observations usable for our purposes. 

Our method in fitting the WISE photometry and deriving effective diameters ($D$) and infrared beaming parameters ($\eta$) is described in Al\'i-Lagoa et al. (\cite{Ali2013}), but with an important caveat explained below. We used the NEATM of Harris (\cite{Harris1998}) and followed the implementation described in Mainzer et al. (\cite{Mainzer2011b, Mainzer2011c}), Masiero et al. (\cite{Masiero2011}), and Grav et al. (\cite{Grav2012}). As discussed in Appendix A of Al\'i-Lagoa et al. (\cite{Ali2013}), our results agree with the NEOWISE team's within the typical error bars of the modelling. 

The main difference here from the method in Al\'i-Lagoa et al. (\cite{Ali2013}) is that we do not use W2 data to avoid a potential bias in the beaming parameter determination related to the assumption that the infrared albedos at W1 and W2 are equal ($p_{W1}=p_{W2}$). This was not problematic in the particular case of the B-types featured in Al\'i-Lagoa et al. because their average IR spectral slope, albeit ranging from slightly blue to moderately red (de Le\'on et al. \cite{deLeon2012}), is small, i.e. they are more neutral. 
When $p_{W1}$ and $p_{W2}$ are significantly different, however, said assumption causes the thermal part of the W2 flux to be miscalculated and introduces a small bias in the best-fitting values of $D$ and $\eta$. Therefore, we only fit the beaming parameter if both W3 and W4 data are available and assume $\eta=1$.0 otherwise\footnote{This is chosen based on the value distributions of all main belt asteroids reported in Masiero et al. (\cite{Masiero2011}).}. We fit $D$ in all cases and the corresponding visible albedos, $p_V$, are derived by means of the relation
\begin{equation}
  p_V = \left(1329\,\left[\mathrm{km}\right] \frac{10^{-H/5}}{D\,\left[\mathrm{km}\right]}\right)^2,
  \label{ec:pVHD}
\end{equation}
where $H$ is the asteroid absolute magnitue taken from the IAU Minor Planet Center (MPC) website. 

Finally, the infrared albedo in filter W1 ($p_{W1}$), is not fitted but computed from the following expression that follows the notation of Emery et al. (\cite{Emery2007}):
\begin{equation}
p_{W1} = \frac{F_{r,W1}r^2\Delta^2}{R^2F_{\odot, W1}\Phi(\theta)},
\label{eq:pIR}
\end{equation}
where ($r$, $\Delta$, $\theta$) is the geometry of observation (the heliocentric distance $r$ must be input in AU), $F_{\odot, W1}$ the solar flux at 1 AU taken from the IR calibration of Rieke et al.~(\cite{Rieke2008}), $F_{r,W1}$ the reflected solar flux measured by the observer at wavelength W1, $R$ the radius, and $\Phi(\theta)$ a phase correction factor (recall that geometric albedo is defined at zero phase angle). We assume that the IAU $H$-$G$ phase correction introduced by Bowell et al.~(\cite{Bowell1989}), namely, 
\begin{equation}
\Phi(\theta) = (1-G)\Phi_1(\theta) + G\Phi_2(\theta),
\label{eq:PHI}
\end{equation}
holds for IR wavelengths. In Eq. \ref{eq:PHI}, $G$ is the so-called slope parameter, and $\Phi_1$ and $\Phi_2$ are empirical functions defined by Bowell et al.~(\cite{Bowell1989}). As $G$ is not determined for the ACOs in our sample, we use the standard assumption $G = 0.15$.  The reflected solar flux $F_{r,W1}$ is estimated by subtracting the model thermal flux at W1 from the observed flux. The former is usually  negligible in comparison to the latter, though not always, in particular, for observations taken when $r\approx1$ AU. 

The assumption $G = 0.15$ may throw off the $H$, and derived albedo values. As shown by Pravec et al. (\cite{Pravec2012}),  $H$ may be underestimated using this $G$-value, in particular for objects with $H > 12$. This implies that ACOs may actually have lower albedos. We  discuss these implications in Sect. 5. In Tables \ref{tabJF1}\addtocounter{table}{1} and  \ref{tabDamocloid1} we list the JF-ACOs and Damocloids, respectively, that we included in this study, and also show the number of measurements from each of the WISE filters that we used in our analysis. 

%In Tabs.  \ref{tabJF1}\addtocounter{table}{1} and  \ref{tabDamocloid1} the number of data fluxes from each of the WISE filters used in the NEATM model are presented along with the absolute magnitudes, the slope parameters, and the observational circumstances of the ACOs in Jupiter Family comets orbits ($JF-ACOs$) and Damocloids studied here.

%%%%%%%%%%%%%%%%%%%%%%%%%%%
%%%%%%%%%%%%%%%%%%%%%%%%%%%
\section{Results and analysis \label{sec:results}}
%%%%%%%%%%%%%%%%%%%%%%%%%%%
%%%%%%%%%%%%%%%%%%%%%%%%%%%

The results of the thermal fits and albedo calculations are summarized in Tables \ref{tabJF2}\addtocounter{table}{1} and \ref{tabDamocloid2}. We obtained $D$ and $p_V$ for 49 of the 203 JF-ACOs and 16 of the 56 Damocloids in Tancredi's list (Tancredi \cite{Tancredi2014}). This is the first published albedo determination for 34 of these 65 ACOs. For 32 of these JF-ACOs and 13 Damocloids, we also obtained $\eta$. We also were able to constrain the infrared-to-visible albedo ratio, $R_{\mathrm{W1}}$, i.e. $p_{W1}/p_V$, for four JF-ACOs and one Damocloid; unfortunately, there was not much reliable photometry in W1 and W2. This is to be expected for small, low-albedo objects such as ACOÕs, which have been observed by WISE at typically large heliocentric distances, as these bands are dominated by reflected light. This small number of $R_{\mathrm{W1}}$ values inhibit our ability to perform any statistical analysis similar to that in the case of B-type asteroids (see Ali-Lagoa et al. \cite{Ali2013}).  %If we consider the best reliable fits corresponds to those objects with more than five observation in filter W3 and W4, only 18 JF-ACOs and 8 Damocloids verify this criterion.

Some of the objects in our survey have been studied previously by Fern\'andez et al. (\cite{Fernandez05}) and by Kim et al. (\cite{Yoonyoung2014}). In fact, Kim et al.  analyzed some of the same WISE data. For the two objects studied by Fern\'andez et al. and the 24 objects studied by Kim et al., which we include  in our analysis, our results are in agreement with these earlier works.

%Using Earth-based mid-IR observations Fern\'andez et al. (\cite{Fernandez05}) computed the albedos of 26 ACOs selected with criteria soften than those in Tancredi (\cite{Tancredi2014}), two of them are presently classified as comets, of the remaining 24 objects 12 are in Tancredi's list (see Table \ref{tabFernandez}). Two of the ACOs in Fern\'andez et al. (\cite{Fernandez05}) were also observed by WISE, (301964) 2000 EJ$_{37}$ and (32511) 2001 NX$_{17 }$, and in both cases the determined diameters are similar within the uncertainties.

%Kim et al.. (\cite{Yoonyoung2014}) present a compilation of the diameter and albedo of 123 ACOs obtained from different published databases using their own selection criteria of ACOs. Only 29 of these objects are in Tancredi's list (see Table  \ref{tabYoun}) and in all but one case (944 Hidalgo) they use the albedo reported by Mainzer et al. (\cite{Mainzer2011d}) using WISE data. In this paper we present our own $D$, $p_V$  and $\eta$ determination of 24 of them, the data of the other 5 did not fulfill the quality we require in Section 2. 
%The albedo determinations in this paper significantly enlarge the number of objects in Tancredi's list with determined size, albedo and beaming parameter. 

%%%%%%%%%%%%%%%%%%%%%%%%%%%
\subsection{Analysis of $p_V$ vs. $T_J$}
%%%%%%%%%%%%%%%%%%%%%%%%%%%

\begin{figure}[h]
  \centering
  \includegraphics[width=8.5cm]{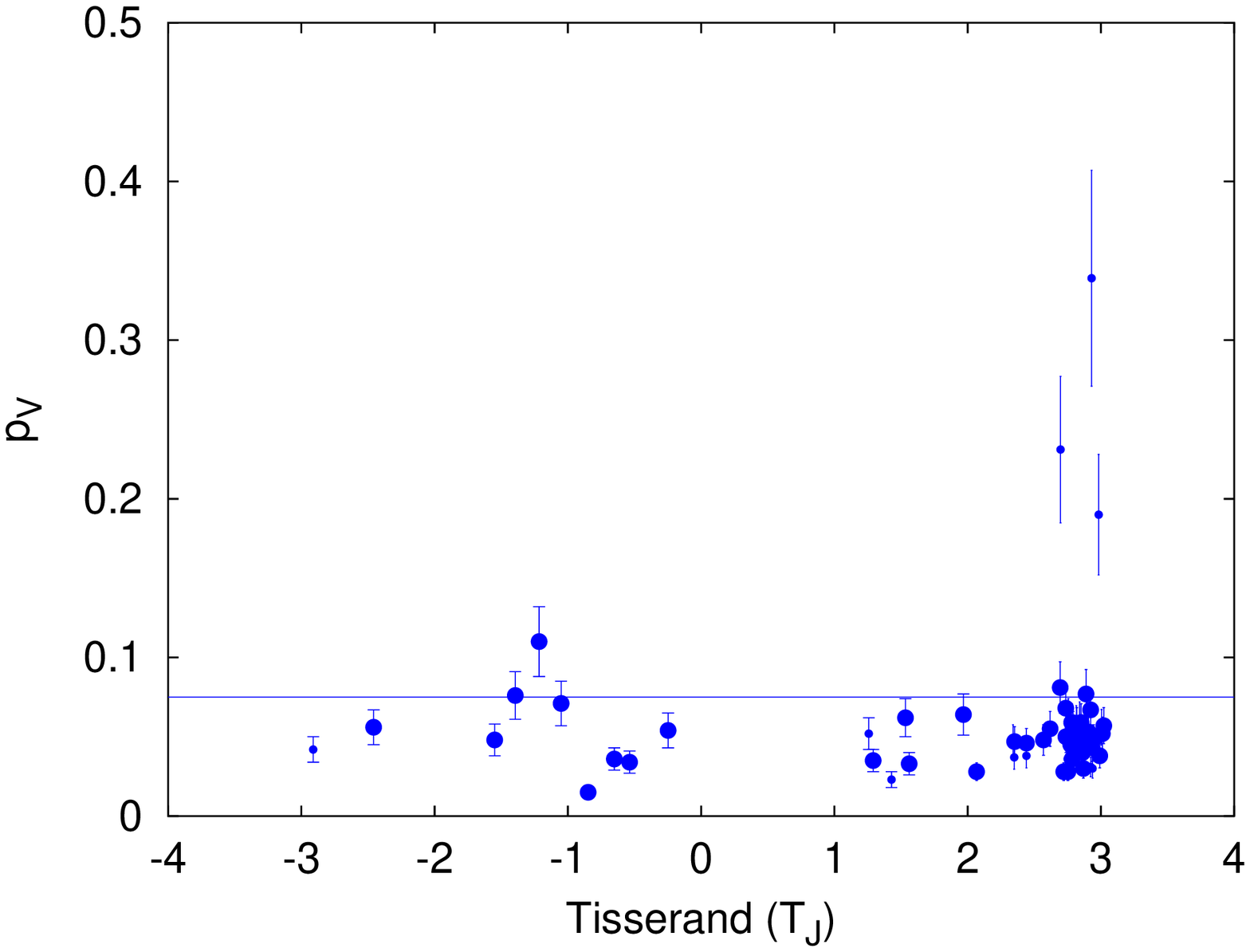}\\
  \caption{\label{figure_vsT}Determined geometric albedo $p_V$ of ACOs vs. Jupiter Tisserand's parameter $T_J$. Big dots correspond to objets with determined $\eta,$ while small dots correspond to ACOs with no $\eta$ determination.  The line at $p_V = 0.075$ is the limit for a cometary-like albedo, according to Fern\'andez et al. (\cite{Fernandez05}). Only three JF-ACOs present a $p_V > 0.18$ incompatible with a cometary origin. }
\end{figure}

Figure~\ref{figure_vsT} presents the geometric albedo against the Tisserand parameter for all objects in our study. All but three objects present an albedo compatible with a primitive object. The JF-ACOs 2005 NA$_{56}$, 2010 KG$_{43}$, and 2010 DD$_2$ have $p_V > 0.18$, which is incompatible with a cometary origin and suggests that they may be scattered igneous asteroids. However , the data of these three asteroids are of lower quantity and quality; we were not able to determine $\eta$ because they only have useful data in filter $W3$. On the other hand, only one Damocloid presents an albedo that is larger than a cometary-like albedo ($p_V \le 0.075$ , according to Fern\'andez et al. \cite{Fernandez05}). The albedo of 2010 OR$_1$, $p_V = 0.110 \pm 0.022$, is in any case a low albedo typical of primitive asteroids. Thus, 95\% of the ACOs in this paper, 61 out of 65, are compatible with a cometary origin, and also not completely incompatible with the albedos of primitive class asteroids (Mainzer et al. \cite{Mainzer2011c}). If we restrict ourselves to the subset of the 65 objects that have enough data to fit $\eta,$ 45 objects, and consider an error bar up to 2-$\sigma$, 
100\% of the sample has a cometary-like albedo.

\begin{table*}
\renewcommand\thetable{2}
\caption{Observational circumstances of the Damocloids included. We also include each object's absolute magnitude $H$, assumed slope parameter $G$, and Tisserand's parameter $T_J$. Columns ``W$n$" ($n=$1 to 4) list the number of useful images in that particular WISE band. Final columns show the heliocentric distance $r$ in AU, distance of the asteroid to WISE spacecraft $\Delta$ in AU, and phase angle $\theta$ in degrees at the midpoint time of the observation. \label{tabDamocloid1}}
\vskip4mm
\centering
\begin{tabular}{lcccccccccc}
\hline
Object                  &$H$    &$G$  &$T_J$  &W1 &W2 &W3 &W4 &$r$  &$\Delta$ &$\theta$\\
\hline
(330759) 2008 SO$_{218}$&12.80 &0.15  &-1.395 &0 &0 &8 &8 &3.687   &4.049  &14.083\\
(336756) 2010 NV$_{1}$  &10.50 &0.15  &-2.911 &0 &0 &1 &4 &9.468   &8.644   &3.775\\
(342842) 2008 YB$_{3}$  &9.30  &0.15  &-0.248 &7 &1 &12 &12 &6.643   &7.454   &4.799\\
2009 QY$_{6}$           &14.80 &0.15  &-0.848  &0 &2 &10 &10 &3.042   &3.591  &14.731\\
2009 YS$_{6}$           &14.20 &0.15  &-1.051  &0 &0 &6 &5 &3.778   &3.343  &14.474\\
2010 BK$_{118}$         &10.20 &0.15  & -2.458 &0 &0 &8 &8 &8.225   &9.016   &3.908\\
2010 CG$_{55}$          &14.20 &0.15  &-1.549 &0 &0 &6 &6 &3.820   &4.156  &13.319\\
2010 CR$_{140}$         &15.50 &0.15  &1.428  &1 &0 &12 &2 &4.197   &4.044  &13.608\\
2010 EJ$_{104}$         &17.10 &0.15  &1.561  &0 &0 &11 &9 &2.315   &2.001  &25.898\\
2010 FH$_{92}$          &11.40 &0.15  &1.533  &0 &0 &9 &9 &5.786   &5.695   &9.901\\
2010 GW$_{64}$          &14.90 &0.15  &-0.537  &0 &0 &9 &6 &3.703   &4.077  &13.729\\
2010 JH$_{124}$         &14.60 &0.15  &1.258  &0 &0 &8 &0 &4.127   &3.923  &14.126\\
2010 JC$_{147}$         &14.70 &0.15  &1.968  &0 &0 &8 &6 &3.452   &3.205  &16.954\\
2010 OR$_{1}$           &16.20 &0.15  &-1.216 &0 &6 &8 &8 &2.050   &1.222  &21.433\\
2010 OA$_{101}$         &18.60 &0.15  &1.291  &0 &6 &10 &8 &1.504   &1.109  &42.496\\
2010 OM$_{101}$         &17.00 &0.15  &-0.652  &0 &0 &9 &5 &2.244   &1.821  &26.398\\
\hline
\end{tabular}
\end{table*}

\begin{table*}
\caption{Results of thermal fits and albedo calculations for Damocloids. We list the best-fitting diameter $D$, the calculated geometric albedo $p_V$, the best-fitting beaming parameter $\eta$ (where applicable), and the near-infrared albedo ratio $p_{\mathrm{W1}}$ (where applicable). \label{tabDamocloid2}} 
\vskip4mm
\centering
\begin{tabular}{lcccc}
\hline
Object                         &$D (km)$   &$p_V$           &$\eta$                 &$p_{\mathrm{W1}}$\\      
\hline
(330759) 2008 SO$_{218}$        &13.5 $\pm$ 1.5 &0.076 $\pm$ 0.015 &0.87 $\pm$ 0.10  &--- \\
(336756) 2010 NV$_{1}$          &52.2 $\pm$ 4.5 &0.042 $\pm$ 0.008 &--- &---\\ 
(342842) 2008 YB$_{3}$          &79.0 $\pm$ 3.0 &0.054 $\pm$ 0.011 &0.895 $\pm$ 0.041  &0.136 $\pm$ 0.059  \\ 
2009 QY$_{6}$                   &12.0 $\pm$ 1.0 &0.015 $\pm$ 0.003 &0.832 $\pm$ 0.080  &---\\ 
2009 YS$_{6}$                           &7.6  $\pm$ 1.7  &0.071 $\pm$ 0.014 &0.94 $\pm$ 0.38  &--- \\
2010 BK$_{118}$                         &52.0 $\pm$ 5.3 &0.056 $\pm$ 0.011 &0.82 $\pm$ 0.11  &--- \\
2010 CG$_{55}$                          &9.9  $\pm$ 2.4  &0.048 $\pm$ 0.010 &0.82 $\pm$ 0.29  &--- \\
2010 CR$_{140}$                         &7.0  $\pm$ 0.4  &0.023 $\pm$ 0.005 &--- &---\\ 
2010 EJ$_{104}$                         &2.8  $\pm$ 0.3  &0.033 $\pm$ 0.007 &1.05 $\pm$ 0.26  &---\\ 
2010 FH$_{92}$                  &28.2 $\pm$ 2.1 &0.062 $\pm$ 0.012 &0.742 $\pm$ 0.085  &---\\ 
2010 GW$_{64}$                          &7.8  $\pm$ 1.3  &0.034 $\pm$ 0.007 &0.86 $\pm$ 0.22  &--- \\
2010 JH$_{124}$                         &7.0  $\pm$ 0.6  &0.052 $\pm$ 0.010 &--- &---\\ 
2010 JC$_{147}$                         &6.3  $\pm$ 1.0  &0.064 $\pm$ 0.013 &1.06 $\pm$ 0.26  &---\\ 
2010 OR$_{1}$                   &2.3  $\pm$ 0.2  &0.110 $\pm$ 0.022 &0.99 $\pm$ 0.19  &---\\ 
2010 OA$_{101}$                         &1.4  $\pm$ 0.3  &0.035 $\pm$ 0.007 &1.18 $\pm$ 0.54  &--- \\
2010 OM$_{101}$                         &3.2  $\pm$ 0.9  &0.036 $\pm$ 0.007 &1.21 $\pm$ 0.58  &---\\ 
\hline
\end{tabular}
\end{table*}

%%%%%%%%%%%%%%%%%%%%%%%%%%%
\subsection{Geometric albedo and beaming parameter distributions}
%%%%%%%%%%%%%%%%%%%%%%%%%%%

\begin{figure}[h]
   \centering
     \includegraphics[width=8.5cm]{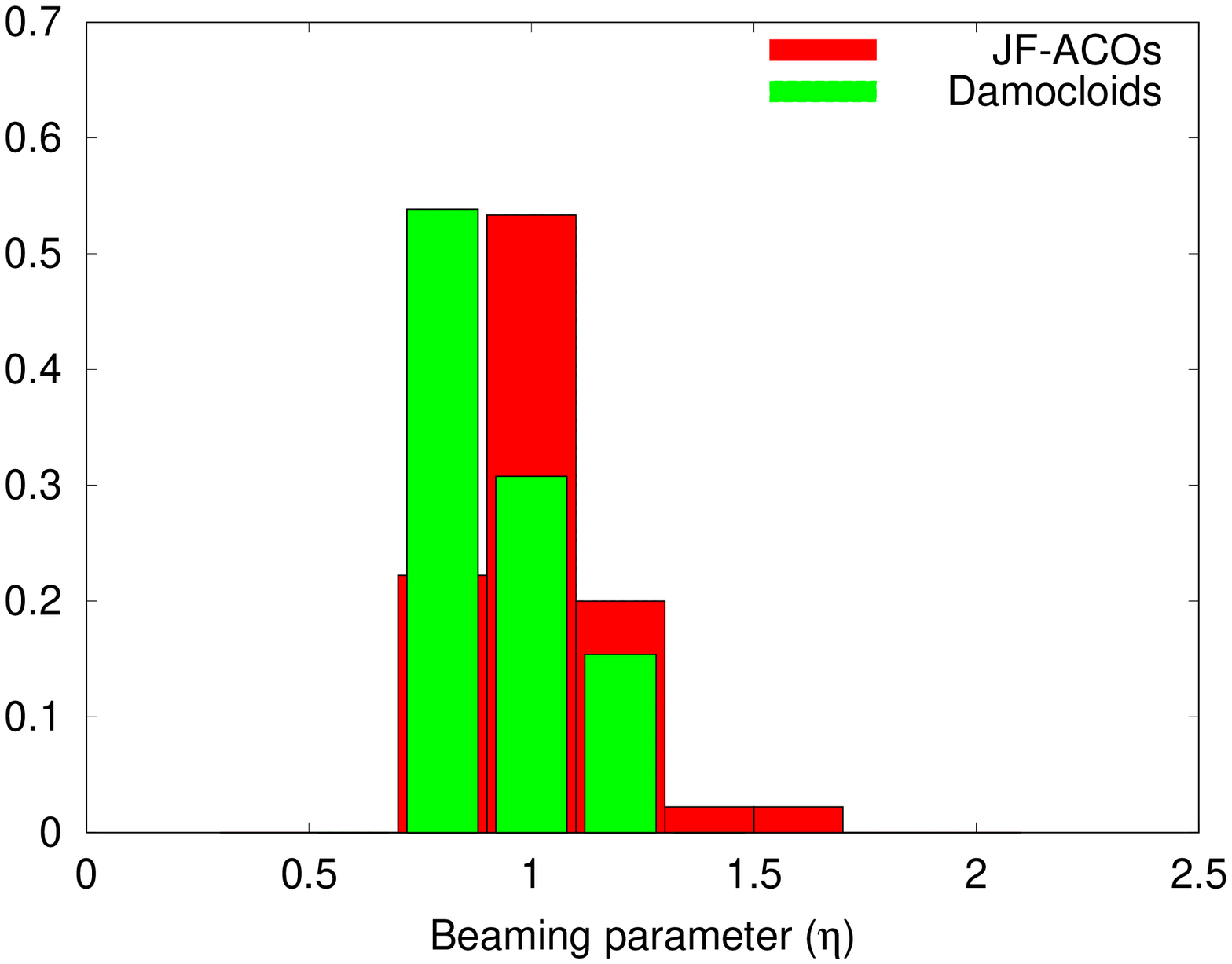}\\
     \includegraphics[width=8.5cm]{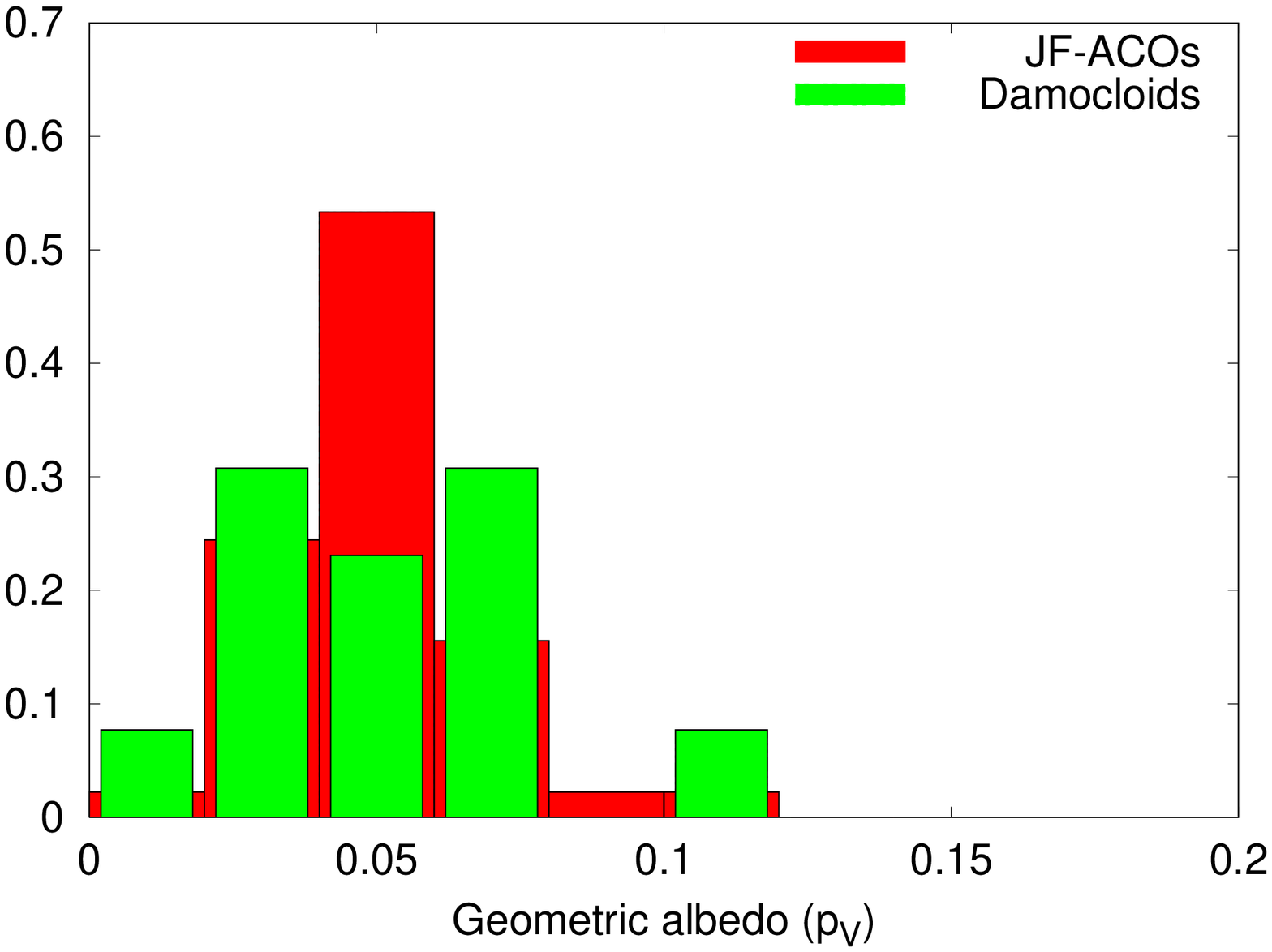}\\
%      \caption{hist_etaACOdicK12, hist_etaACOdicK12w3w4gt5}
         \caption{\label{figure_hist} {\em Upper:} beaming parameter distribution of JF-ACOs (red) and Damocloids (green) presented as normalized histograms; {\em Lower:} geometric albedo distribution of JF-ACOs (red) and Damocloids (green) with determined $\eta$.}
   \end{figure}

The normalized histograms of  Fig. \ref{figure_hist} represent the beaming parameter (upper) and geometric albedo (lower) distributions of JF-ACOs and Damocloids with determined $\eta$ (those with better WISE data) are presented.
% excluding the 3 JF-ACOs mentioned earlier with $p_V > 0.18$ that are incompatible with a cometary origin.

The albedo distributions of Damocloids and JF-ACOs are similar and narrow around the mean value, which indicates that both populations have similar surface properties. The mean geometric albedo (and 1$\sigma$ standard deviation of the mean) for JF-ACOs is  $\bar{p_V} = 0.048 \pm 0.013$, whereas for Damocloids it is $\bar{p_V} =0.051 \pm 0.023$. If we assume that both distributions are compatible, the Kolmogorov-Smirnov test (K-S hereafter) cannot reject this null hypothesis with a high $P$-value of 0.67. Thus, for the entire sample of ACOs the mean geometric albedo  is $\bar{p_V} = 0.049 \pm 0.016$.

The $\eta$ distributions of Damocloids and JF-ACOs look slightly different but are compatible within their error bars, with a mean beaming parameter $\bar{\eta} =0.94 \pm 0.14$ for the Damocloids and $\bar{\eta} =1.05 \pm 0.16$ for the JF-ACOs. The K-S test gives us a borderline case\footnote{Null hypothesis rejection criteria range from the less strict $P < 0.05$ to the more strict $P < 0.03$ or even $P < 0.01$} with a $P$-value of 0.04. The reason for this is likely the presence of four JF-ACOs, which have somewhat elevated $\eta \sim 1.21 -- 1.6$, but given the large error bars in $\eta$, we consider that both distributions are not incompatible at the $P\leq 0.01$ level. Larger samples would be needed to be absolutely conclusive and it would also be interesting to have more data to confirm those objects with high $\eta$-values.

As we show later, the small differences in the $\bar{\eta}$-values could be related to the fact that Damocloids were observed at slightly lower phase angles (mean phase angle $\bar{\theta} = 17.1$\degree\ versus $\bar{\theta} = 21.1$\degree\ for the JF-ACOs). The only object with an $\eta$ well above 1.0 is 2008 UD$_{253}$, with $\eta =1.6 \pm 0.7$, but  the uncertainty is large and so is $<1\sigma$ away from 1.0. If we had used a fixed $\eta = 1.0$ in the best-fitting size for the object would still result in a cometary-like albedo of $p_V = 0.047$. For the entire sample of ACOs, the mean beaming parameter is $\bar{\eta} =1.02 \pm 0.16$. This is compatible with the $\bar{\eta} =1.0$ of the main belt asteroids measured with WISE (Masiero et al. \cite{Masiero2011}). On the other hand, Delbo et al. (2007) found a $\bar{\eta} =1.5$ for near-Earth asteroids (NEAs), and Mainzer et al. (\cite{Mainzer2011d}) report a median $\eta = 1.4 \pm 0.5$ for 313 NEAs. Our results are far from these values, although we emphasize that only two of our JF-ACOs are actually NEAs, $2009 WF_{104}$ and $2009 WO_6$, whose $\eta$-values, 1.15 and 1.16, are still closer to $\bar{\eta} =1.0$. Also, that  a large fraction of NEAs in Delbo et al. and Mainzer et al. are S-types, which have different ensemble thermal properties than the primitive asteroids.

Even more relevant is to compare these distributions to those of comets. Fern\'andez et al. (\cite{Fernandez13}) presented mid-infrared measurements of the thermal emission from 89 nuclei of Jupiter family comets (JFCs) obtained with the {\em Spitzer} Space Telescope, and they were able to measure $\eta$ for 57 of those nuclei.  They derived the weighted mean and standard error on the mean of the $\eta$ of the JFC population $\bar{\eta} =1.03 \pm 0.11$, which is very similar to the distribution we found for ACOs. 

The comparison with the albedo distribution of JFCs is not straightforward; the albedo is computed using the diameter and the absolute magnitude, and the absolute magnitudes of comet nuclei are usually quite uncertain. It is a difficult task to measure them because a nucleus is usually embedded in a dust coma (see e.g. Licandro et al. \cite{Licandro2000}).  By comparing the size distribution determined with {\em Spitzer} data and those determined using visible data, Fern\'andez et al. (\cite{Fernandez13})  conclude that ``a geometric albedo of 0.04 is a reasonable assumption for visible-wavelength studies" in agreement with previous results by Lamy et al. (\cite{Lamy2004}), which imply a narrow albedo distribution around a $\bar{p_V} \sim 0.04$. 

To test if this is compatible with the albedo distribution of ACOs found here, we constructed a synthetic set of albedo values as follows. We took $D_\mu = 10$ km and $H_\mu = 14.10$ so that the corresponding visible geometric albedo is $p_V \sim 0.04$ (per Eq.~1). Then, we generated 60 random values of size and absolute magnitude, each one drawn from Gaussian distributions with mean values $D_\mu$ and $H_\mu$ and standard deviations of 1.5 km (15\% relative error in size) and 0.15 magnitudes, respectively. The resulting $p_V$ distribution represents what we would obtain if we computed or measured the size and absolute magnitude for a single object with these levels of uncertainty, which we consider a plausible representation for a population of objects with the same visible geometric albedo. We used this $p_V$ distribution as the null hypothesis for a K-S test, and we obtained that it is not compatible with the albedo distribution of the ACOs. This means that, however small, there is a statistically significant shift between the assumed cometary nuclei $p_V = 0.04$ and the ACOs visible geometric albedos. On the other hand, the null hypothesis was not rejected when we decreased $H$ to 13.85, which corresponds to a slightly higher albedo value of 0.045, which in any case,  is very close to the assumed mean cometary albedo. From this, we conclude that, accounting for our typical uncertainties, the $p_V$ distribution of the ACOs is similar to the $p_V$ distribution of comet nuclei: very narrow around a small typical geometric albedo $p_V = 0.045$, an albedo value slightly larger, but still compatible with that assumed by Lamy et al. (\cite{Lamy2004}) and Fern\'andez et al. (\cite{Fernandez13}) for comet nuclei.

%%%%%%%%%%%%%%%%%%%%%%%%%%%
\subsection{Analysis of $p_V$ and $\eta$ versus Diameter}
%%%%%%%%%%%%%%%%%%%%%%%%%%%

\begin{figure}[h]
   \centering
     \includegraphics[width=8.5cm]{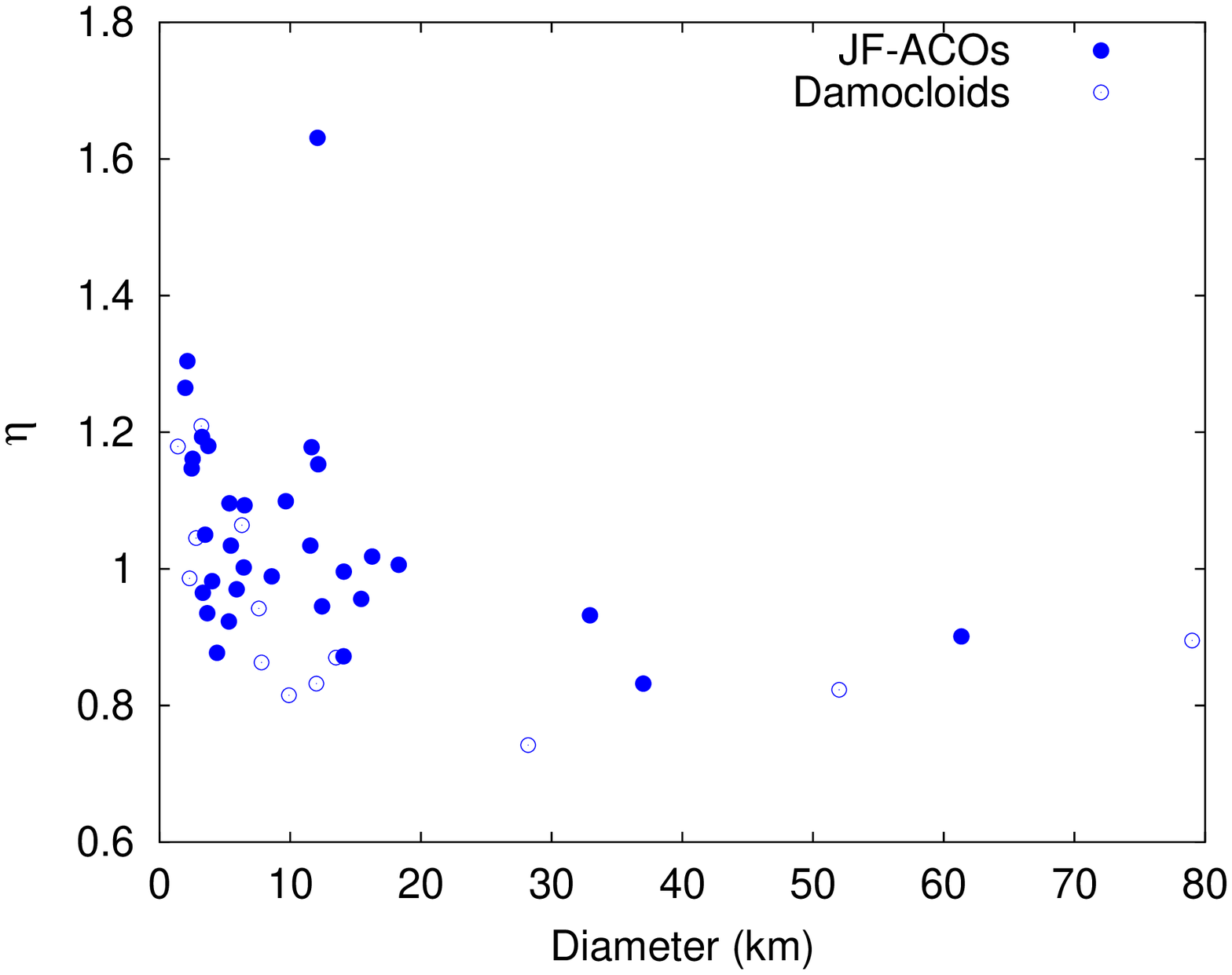}\\
     \includegraphics[width=8.5cm]{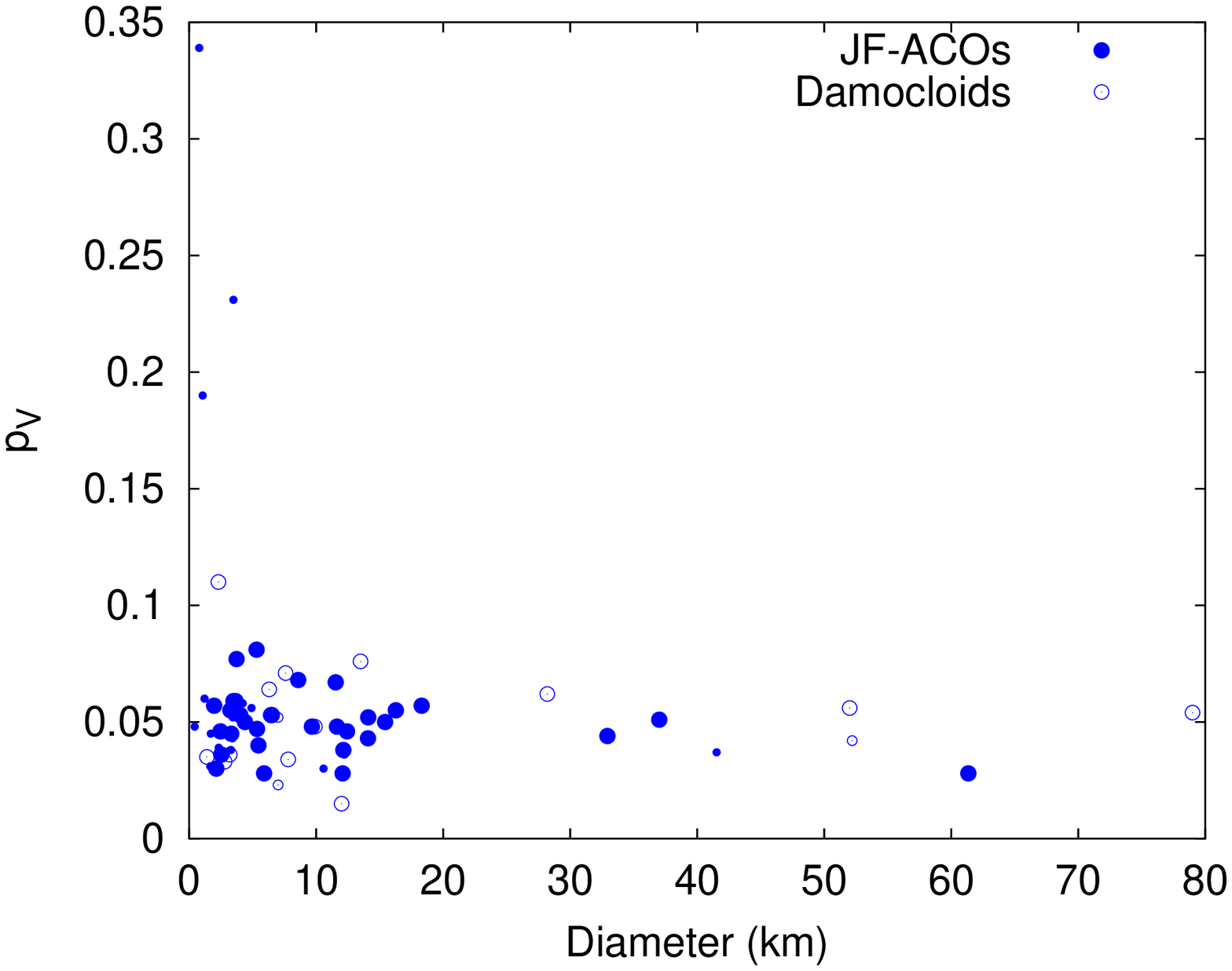}\\
%      \caption{hist_etaACOdicK12, hist_etaACOdicK12w3w4gt5}
           \caption{{\em Upper:} beaming parameter vs. diameter distribution of JF-ACOs (filled circles) and Damocloids (open circles). A possible trend of $\eta$ with $D$ is shown. {\em Lower:} geometric albedo vs. diameter distribution of JF-ACOs  and Damocloids.  Large circles correspond to objects with  $\eta$ determined and small objects with not enough data to determine $\eta$. No correlation between $p_V$ and $D$ is found.
             \label{figure_Diameter}}
   \end{figure}

We also looked for possible correlations of $p_V$ and $\eta$ with $D$. Fig. \ref{figure_Diameter} presents these scatter plots. First, we find there is no correlation between the albedo and the diameter, which is similar to the case of JFCs according to Fern\'andez et al. (\cite{Fernandez13}). Hence, it is reasonable to assume a common geometric albedo for all JFCs and JF-ACOs for statistical purposes, e.g. to compute their size distribution from their visible absolute magnitude, something that is frequently invoked when interpreting visible-wavelength photometry of JFCs. 

Also, the three objects with high albedo values have very small diameters ($D < 3.5 km$), which may be the reason they have lower-SNR WISE data and thus poorly-constrained $D$ and $p_V$. On the other hand, these values of the diameter suggest that the Yarkovsky effect could have played an important role in their being scattered from the main belt.

A possible trend of $\eta$ with $D$ is apparent in the upper panel of Fig. \ref{figure_Diameter}. In  Sect. \ref{sec:resultsvsalpha} we show that this is likely due to a phase angle effect.

%%%%%%%%%%%%%%%%%%%%%%%%%%%
\subsection{Analysis of $p_V$ and $\eta$ vs. the phase angle of the observations}\label{sec:resultsvsalpha}
%%%%%%%%%%%%%%%%%%%%%%%%%%%

\begin{figure}[h]
   \centering
     \includegraphics[width=8.5cm]{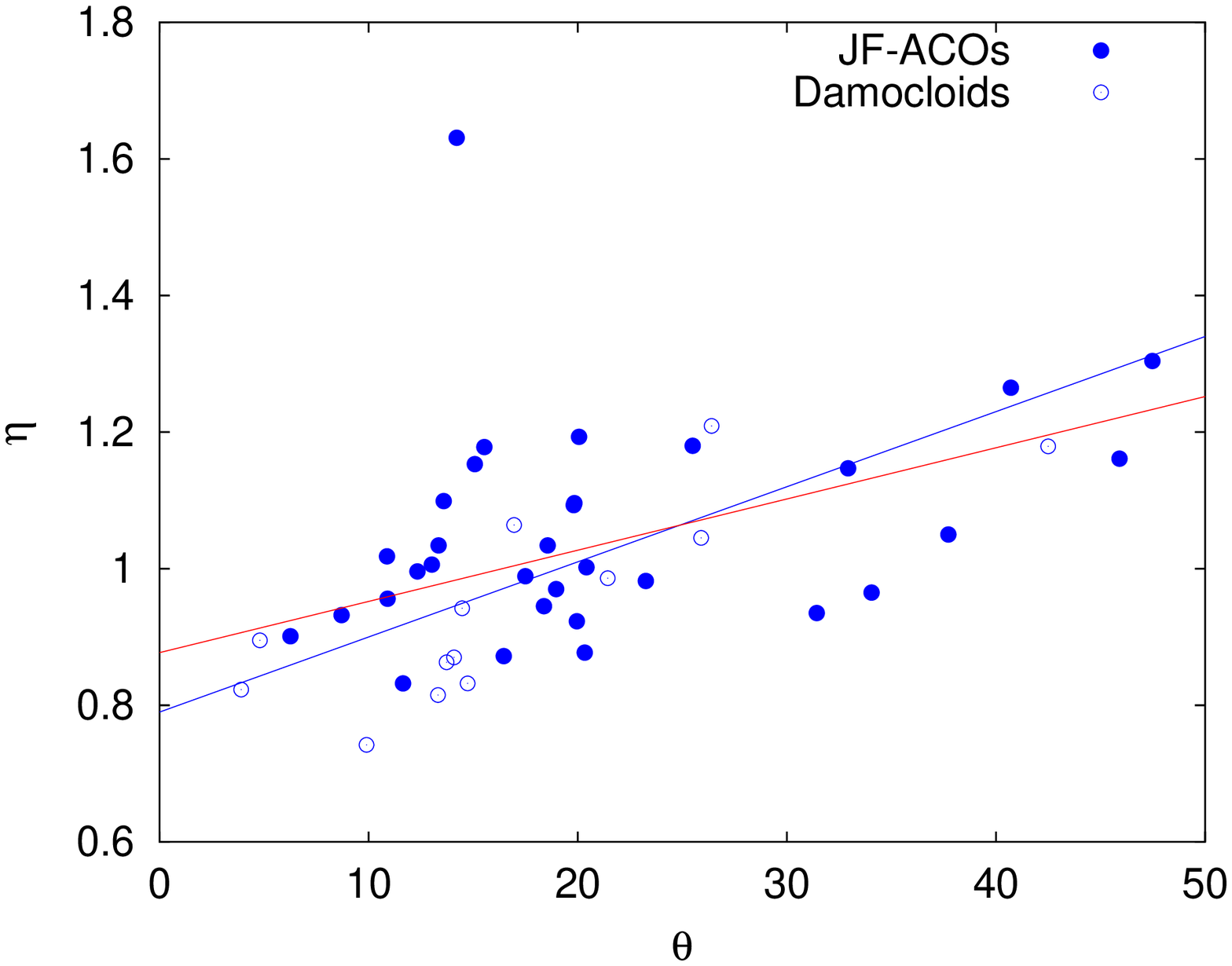}\\
     \includegraphics[width=8.5cm]{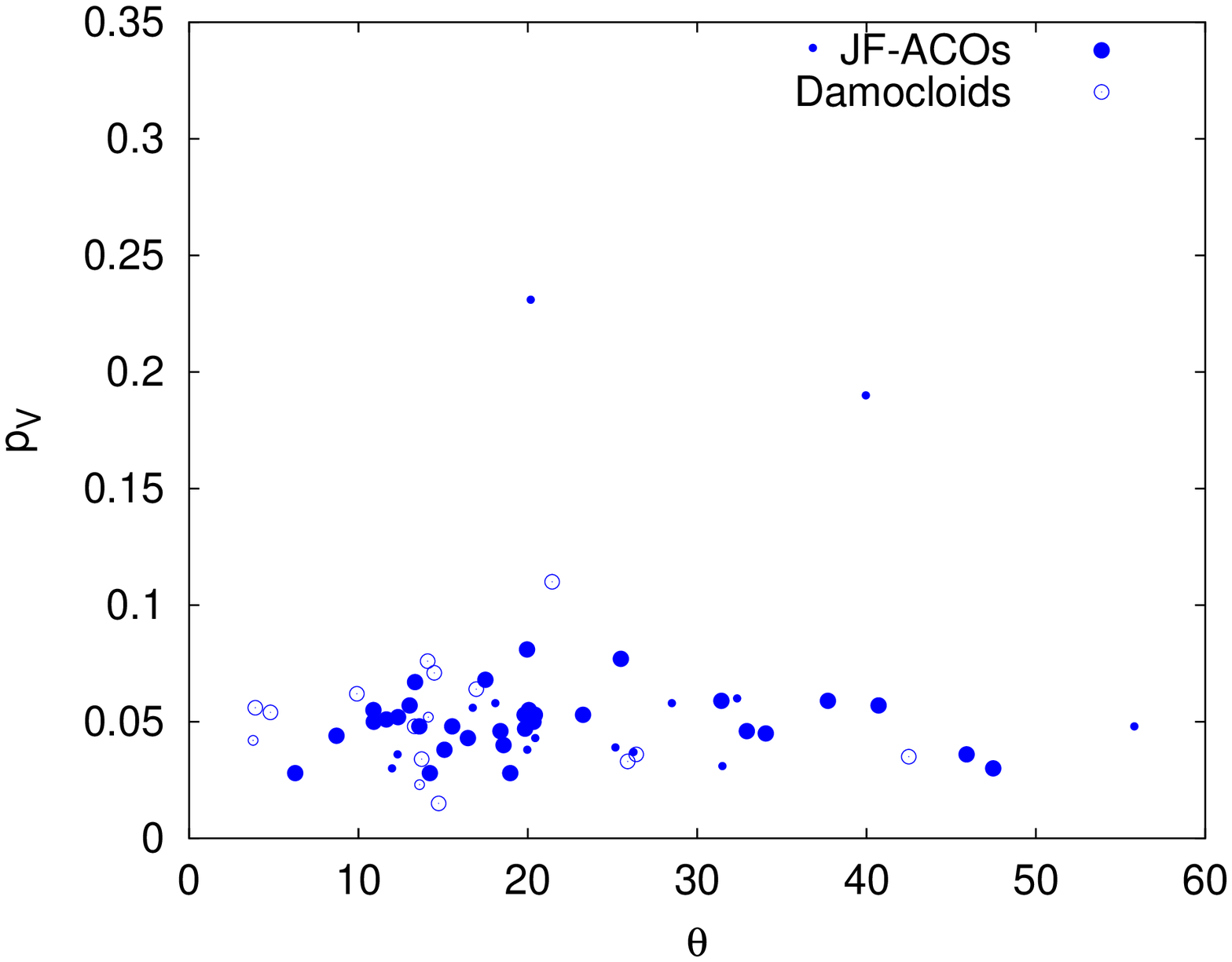}\\
%      \caption{hist_etaACOdicK12, hist_etaACOdicK12w3w4gt5}
     \caption{{\em Upper:} beaming parameter vs. phase angle  of JF-ACOs (filled circles) and Damocloids (open circles). A linear  trend of $\eta$ with $\theta$ is shown (see text); the red line is the correlation we found  ($\eta = 0.85 + 0.008 \  \theta$), the blue line is the linear regression ($\eta = 0.79 + 0.011 \  \theta$) from Masiero et al. (\cite{Masiero2011}). {\em Lower:} geometric albedo vs.  $\theta$ distribution of JF-ACOs  and Damocloids. Large circles correspond to objects with  $\eta$ determined and small objects with not enough data to determine $\eta$. No correlation between $p_V$ and  $\theta$ is found. \label{figure_hist3}}
   \end{figure}

We also searched for a correlation between geometric albedo and phase angle, and $\eta$ with phase angle. Figure \ref{figure_hist3}  shows $p_V$ vs. $\theta$ (lower plot) and $\eta$ vs. $\theta$ (upper plot). No correlation is found between albedo and phase angle, but a correlation between beaming parameter and phase angle is evident. We find a best-fit linear relation of beaming parameter as a function of phase of $\eta(\theta) = (0.85 \pm 0.03) + (0.008 \pm 0.001) \  \theta$. This result is similar to that presented in Masiero et al. (\cite{Masiero2011}): $\eta(\theta) = 0.79 + 0.011 \  \theta$.

\begin{figure}[h]
   \centering
     \includegraphics[width=8.5cm]{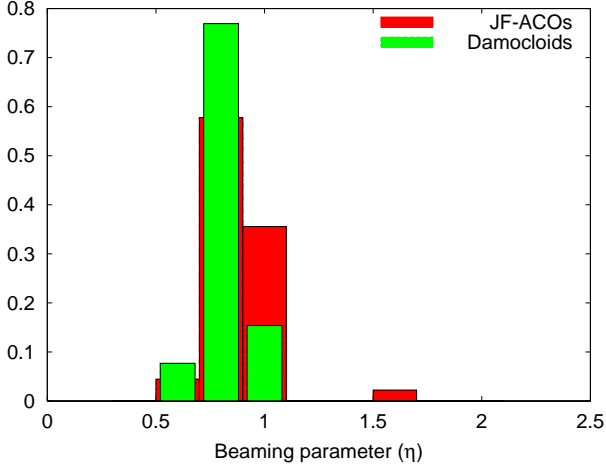}\\
     \caption{Normalized histogram of $\eta$ corrected to $\theta = 0$ using the linear fit obtained in this paper.  Both distributions are very similar.
       \label{figure_hist_corr}}
   \end{figure}

\begin{figure}[h]
   \centering
     \includegraphics[width=8.5cm]{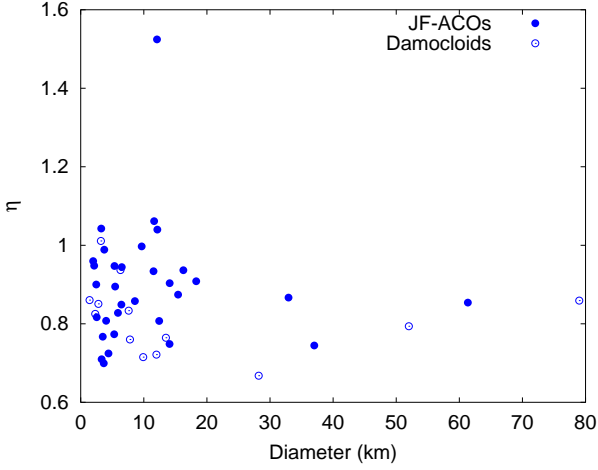}\\
     \caption{Beaming parameter corrected to $\theta = 0$ using the linear fit obtained in this paper vs. diameter. The apparent trend of $\eta$ with $D$ in Fig. 3 is not anymore observed in this plot. \label{figure_etaD_corr}}
   \end{figure}

   As discussed above, there is a difference between the JF-ACO and Damocloid $\eta$ distributions, and a possible correlation between size and $\eta$ (see Fig.~\ref{figure_Diameter}). Once we correct the $\eta$-values as a function of $\theta,$ using the linear fit found above, the $\eta$ distributions become even more similar (see Fig. \ref{figure_hist_corr}).  The K-S test distance is reduced and the $P$-value is 0.065, i.e. the distributions are not incompatible. Furthermore, after applying this correction to $\eta,$ we found that the possible correlation of $\eta$ with $D$ also disappears (see Fig.~\ref{figure_etaD_corr}).

Fern\'andez et al. (\cite{Fernandez13}) found that the beaming parameter of comet nuclei is uncorrelated with any observational or geometric parameter; in particular, they did not find a correlation with the phase angle, although their observations only covered  a narrow range of phase angles ($ 9 < \theta < 17.5 \degree$). It is likely that this was too narrow a range to see the kind of correlation we report here for the ACOs, if it exists for the comet nuclei.

%%%%%%%%%%%%%%%%%%%%%%%%%%%
%%%%%%%%%%%%%%%%%%%%%%%%%%%
\section{ACOs cumulative size distribution \label{sec:sizes}}
%%%%%%%%%%%%%%%%%%%%%%%%%%%
%%%%%%%%%%%%%%%%%%%%%%%%%%%

\begin{figure}[h]
 \centering
   \includegraphics[width=8.5cm]{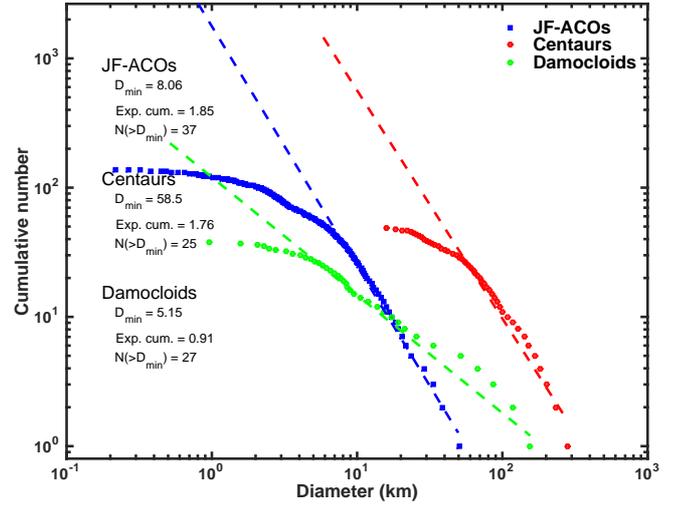}\\
    \caption{Cumulative size distribution for JF-ACOs, Centaurs, and Damocloids (see text).     \label{figure_CSD_ACOS}}
\end{figure}

Alvarez-Candal \& Licandro~(\cite{AlvarezCandal2006}) studied the CSD of the population of JF-ACOs and derived a CSD power-law index of  $\beta = 2.55 \pm 0.04$. However, they used a sample of objects that was strongly contaminated by scattered asteroids because of their selection criteria. 

We now present CSDs for the three groups of ACOs: Jupiter family, Damocloids, and Centaurs. The sample of objects in each group with reliable size estimates is rather small: 49 JF-ACOs, 16 Centaurs and 16 Damocloids. In the case of Centaurs, we used the results for the 16 objects presented in Bauer et al. \cite{Bauer2013} that are in Tancredi's list. These sample sizes are not sufficient by themselves however to find reliable CSDs. We follow the procedure presented by Tancredi et al.~(\cite{Tancredi2006}) to enlarge the sample. We start with  the sample of 203 JF-ACOs, 72 Centaurs, 56 Damocloids included in Tancredi's list (\cite{Tancredi2014}). For each individual object, we assume a distribution of the albedo according to available information, to wit: 
\begin{itemize} 
\item If a given object has a reliable measurement of the albedo with a certain error bar, we assume for this object an albedo distribution with a Gaussian shape (within $\pm 2 \sigma$), 
centred on the reported value and with $\sigma$ equal to the reported error. The values for the JF-ACOs and Damocloids are taken from Tables \ref{tabJF2} \& \ref{tabDamocloid2}, and those of the Centaurs from the albedos presented in Bauer et al. (\cite{Bauer2013}). 
\item If we do not have any information about the albedo of the object, we assume an albedo distribution with a normal shape (within $\pm 2 \sigma$ ), centred at the mean and standard deviation of the values for each group.
\end{itemize}

Then, from the probability density function (PDF) of albedo for each individual object, following Eq. 7 of Tancredi et al.~(\cite{Tancredi2006}), we compute the PDFs for the diameter for a given value of the absolute magnitude ($H$). In the case of ACOs, $H$ is equivalent to the absolute nuclear magnitude ($H_N$) used in the case of comets (see Tancredi et al.~\cite{Tancredi2006}). The values of $H$ are taken from the Minor Planet Center. Adding the individual PDF $p_i(\log{D})$, we obtain the differential size distribution (in terms of $\log{D}$) for the whole sample, from which we can easily compute the CSD.

%(PONEMOS UN LISTADO CON LOS VALORES DE H_N QUE USAMOS??)

The CSDs computed with this procedure for each ACO group are presented in Fig. \ref{figure_CSD_ACOS}. The size distributions are fitted to a power-law probability density $p$ of the type
\begin{equation}
p(D) = C D^{-\alpha} \ \ ,
\end{equation}
where $\alpha$ is the constant exponential parameter known as the exponent and $C$ a normalization constant. The power law represents asteroids down to some cut-off diameter ($D_{min}$), giving the normalization constant $ C = (\alpha - 1) D_{min}^{\alpha - 1}$, provided $\alpha > 1$. The cumulative distribution function ($P$) is then
\begin{equation}
P(>D) = \left(\frac{D}{D_{min}}\right)^{-(\alpha-1)} \ \ 
.\end{equation}

The exponents of the distributions ($\alpha$) and the cut-off diameters ($D_{min}$) are estimated with a method, which combines a maximum-likelihood fitting method with goodness-of-fit tests based on the Kolmogorov-Smirnov statistic, as described by Clauset (\cite{Clauset2009}) and Tancredi et al. (\cite{Tancredi2015}) (see an example of the method below). The exponent of the CSDs ($\beta =\alpha-1$), $D_{min}$  and the number of objects larger than $D_{min}$ ($N(>D_{min})$) are presented in Fig. \ref{figure_CSD_ACOS}.

The linear part in the log-log plot of the CSDs for JF-ACOs and Centaurs are almost parallel with a similar cumulative exponent ($\beta = 1.85 \pm 0.30 , = 1.76 \pm 0.35$, respectively). The CSD for Centaurs is shifted towards larger objects with respect to that for JF-ACOs. For example, for $D > 30$ km there are 25 times more Centaurs than JF-ACOs. The exponent for the Centaurs  is similar to that  obtained by Bauer et al. \cite{Bauer2013} ($\beta = 1.6 \pm 0.30$), although there are differences in the samples and computation method.

On the other hand, the CSD for Damocloids presents a much shallow exponent $\beta = 0.89 \pm 0.17$.
In the light of the hypothesis that all the ACOs are dormant comets, we have to bear in mind that the mantling process might alter the original size distribution. This process might be dependent on the size of the nucleus, chemical constituents, and  number of perihelion passages. The time required for the mantling is expected to be much longer for the Damocloids than for the JF-ACOs because of their longer orbital periods. Nevertheless, if the mantling is only dependent on the number of passages, in a steady state scenario, the CSDs should not be affected. Therefore, the differences among the exponents may reflect a difference in the CSDs previous to the mantling process and/or mantling processes that depend on the size of the nucleus in different ways. 

We conclude that the similarity in the exponent between JF-ACOs and Centaurs and the difference with the Damocloids reinforces the hypothesis that the populations of the former two populations have a common origin (in the transneptunian belt), while the latter comes from a different origin (in the Oort Cloud).

\begin{figure}[h]
 \centering
    \includegraphics[width=8.5cm]{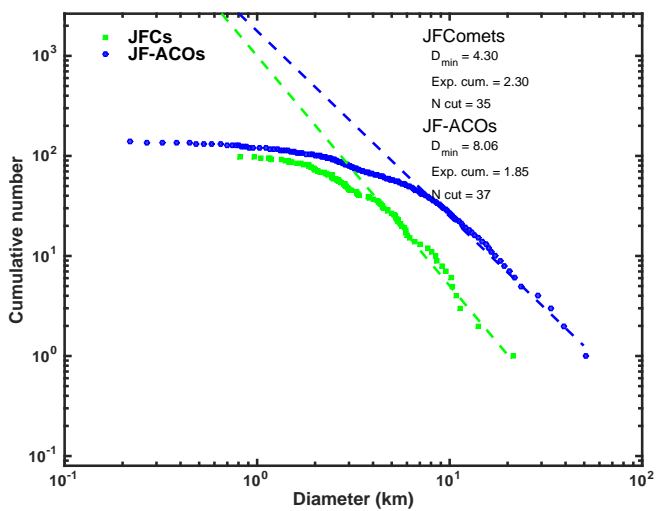}\\
    \caption{Cumulative size distribution  for JF-ACOs and JFCs. \label{figure_CSD_ACOS_JFC}}
\end{figure}

\begin{figure}[h]
 \centering
    \includegraphics[width=8.5cm]{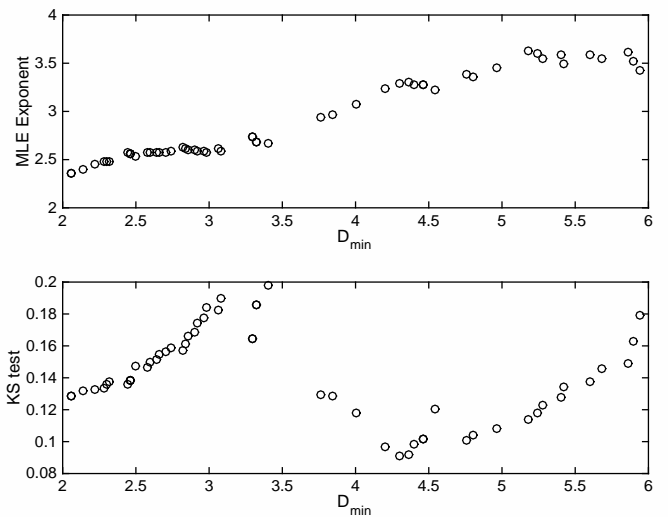}\\
    \caption{Value of the Kolmogorov-Smirnov test as a function of the cut-off diameter ({\it lower plot}) and the corresponding value of the maximum-likelihood exponent $\alpha$ ({\it upper plot}).     \label{figure_KStest_JFC}}
\end{figure}

Now we compare the CSD of JF-ACOs with the sample of active Jupiter family comets (JFCs). The CSDs of the JFCs has been analysed by many authors: {\it e.g.} Lamy et al. (2004); Meech et al. (2004); Tancredi et al. (2006); Weiler et al. (2011); Snodgrass et al. (2011); and Fern\'andez et al. \cite{Fernandez13}. There is no agreement among these authors and there has been a strong debate about the exponent of the CSDs. In our analysis we use the sample of diameters for the JFCs presented by Fern\'andez et al. (\cite{Fernandez13}). The sample includes 89 JFCs from their Table 7, plus nine objects already included in their analysis whose effective radii had been measured by spacecraft flybys or by earlier mid-IR photometry. The CSDs for JF-ACOs and JFCs are presented in Fig. \ref{figure_CSD_ACOS_JFC}. 
The computation of exponents of the distributions and the  cut-off diameters was peformed with the method described above. As an example, in Fig. \ref{figure_KStest_JFC} we present the value of the Kolmogorov-Smirnov statistic as a function of the cut-off diameter ({\it a}) and the corresponding value of the maximum-likelihood exponent $\alpha$ ({\it b}) for the sample of JFCs. The KS-statistic is computed as the maximum distance between the size distribution of the data and the fitted model for each cut-off diameter. A similar plot could be presented for the JF-ACOs. The chosen value corresponds to the minimum KS-statistic  at $D_{min} = 4.30$ km, with a corresponding $\alpha = 3.30$ ($\beta = 2.30 \pm 0.39$). If a smaller cut-off diameter is selected, the value of $\alpha$ is drastically reduced. Fern\'andez et al. (\cite{Fernandez13}) reported an exponent $\beta = 1.92 \pm 0.23$, computed as the median among different linear fits in several range of diameters. Although the two estimates overlap, we  use our value to compare with the other samples for consistency.

From the comparison of the two CSDs (Fig. \ref{figure_CSD_ACOS_JFC}) we conclude that the CSD for JF-ACOs is shallower and shifted towards larger diameter with respect to the JFCs. In the context of the hypothesis that the JF-ACOs are dormant comets coming from the JFCs population, this result has several implications:
\begin{itemize}
\item The largest JF-ACOs are much larger than the largest JFCs; {\it e.g.} there is only one JFC nucleus larger than 20 km, but there are many JF-ACOs larger than this size. 
\item The mantling process has a size dependency; large comets tend to acquire an inactive stage faster than small comets, which is consistent with previous results that suggest that larger comets develop insulating dust mantles faster (Licandro et al. \cite{Licandro2000} , Tancredi et al. \cite{Tancredi2006})
\item The number of JF-ACOs is higher than the JFCs for a given size, although we have to take into account that the sample of measured JFC nuclei is smaller than the known sample. 
\end{itemize}

While for the sample of JF-ACOs, we are using the complete sample of discovered objects, we only use the sample of JFCs with measured absolute magnitudes, which is much smaller than the sample of discovered objects. In order to compare the numbers of the two population, we hence correct the CSDs of JFCs taking  the incompleteness of the sample used to compute the CSD presented in Fig. \ref{figure_CSD_ACOS_JFC} into account. Though we may expect that the incompleteness has some dependency on the nuclear magnitude and the diameter, in the absence of any further information we assume a constant incompleteness factor equal to the ratio of the total number of known JFCs over the sample with measured diameters ($=476/98$).  

\begin{figure}[h]
 \centering
   \includegraphics[width=8.5cm]{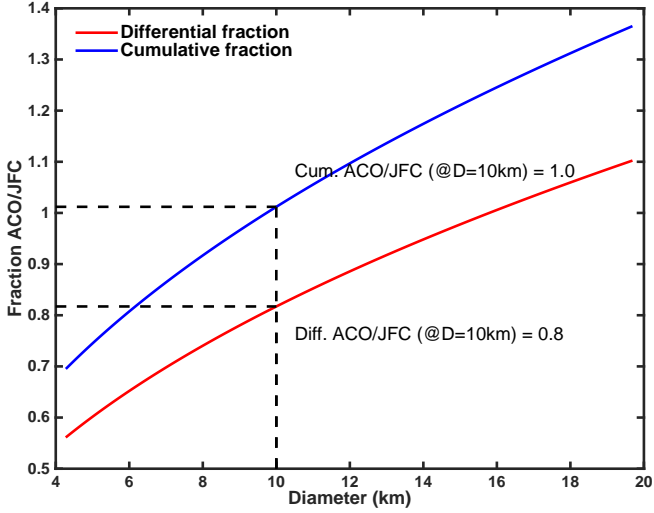}
   \caption{\label{figure_fraction_ACOS_JFC}Ratio between the differential (red line) and the cumulative (blue line) number of JF-ACOs over the number of JFCs (corrected for incompleteness) for a given diameter (see text).}
\end{figure}

In Fig. \ref{figure_fraction_ACOS_JFC} we present the ratio between the differential (red line) and the cumulative (blue line) number of JF-ACOs over the number of JFCs (corrected for incompleteness) for a given diameter. These ratios are size-dependent: the ratios increase with increasing diameters. At $D = 10$ km, the differential ratio is 0.8 and the cumulative ratio is 1. Thus, we conclude that the population of JF-ACOs is comparable in number with the population of JFCs. 

This result is in contradiction with previous estimates of a large population of dormant extinct comets, as has already been pointed out in Tancredi (\cite{Tancredi2014}). Shoemaker et al. (\cite{Shoemaker1994}) estimated a ratio of extinct to active JFCs of $\sim$7.5; and Levison and Duncan (\cite{Levison1997}) obtained a ratio of 3.5. Our results contradict these estimates  based on the present-day number of JF-ACOs and JFCs.

Although the cumulative number of JF-ACOs and JFCs larger than a few km seem to be similar, we find that for very large objects ($> 20 km$), there is a prevalence of JF-ACOs over JFCs. There is a size over a few tens-km where only JF-ACOs have been observed. We can tentatively conclude that tens-km rapidly builds a insulating crust and/or they are physically unstable.

%%%%%%%%%%%%%%%%%%%%%%%%%%%
%%%%%%%%%%%%%%%%%%%%%%%%%%%
\section{Conclusions \label{sec:discussion}} 
%%%%%%%%%%%%%%%%%%%%%%%%%%%
%%%%%%%%%%%%%%%%%%%%%%%%%%%

In this paper we present the effective diameter ($D$), infrared beaming parameter $\eta$, and visible geometric albedos ($p_V$) for 49 of the 203 JF-ACOs and 16 of the 56 Damocloids from the Tancredi~\cite{Tancredi2014} list derived from applying the NEATM to their WISE data. The beaming parameter could be fitted for  32 of these JF-ACOs and 13 Damocloids.  

All but three JF-ACOs present an albedo that is compatible with a cometary origin.  The data quality of those three asteroids is relatively worse and that could explain the discrepancy. All Damocloids present an albedo typical of primitive asteroids and compatible with a cometary origin. Taking both populations, the albedo of 95\% of the ACOs of our sample are compatible with a cometary origin. If we restrict our sample to those objects with enough data to fit $\eta$ (45 objects), the result is 100\%. 

We did not find any correlation between: (1) the albedo and the Diameter; (2) the albedo and phase angle of the observation; (3) $\eta$ and the albedo. On the other hand, we find a linear relation between $\eta$  and phase angle $\theta$ ($\eta=0.85 + 0.008 \ \theta$) similar to that presented in Masiero et al. (\cite{Masiero2011}) for the whole main belt data set ($\eta =0.79 + 0.011 \ \theta$). Applying this correction to the computed  $\eta$ values, we conclude that there is no correlation between $\eta$ and $D$. 

The $p_V$ and $\eta$ distributions of Damocloids and JF-ACOs are similar, which is indicative that both populations have similar surface properties. Their albedo distributions are very narrow around their mean values: $\bar{p_V} = 0.048 \pm 0.013$ and $\bar{p_V} =0.051 \pm 0.023$ for JF-ACOs and for  Damocloids, respectively. For the entire sample of ACOs, the mean geometric albedo  is $\bar{p_V} = 0.049 \pm 0.016$, compatible with a narrow albedo distribution around a $\bar{p_V} \sim 0.05$ similar to the narrow albedo distribution around a $\bar{p_V} \sim 0.04$ assumed by Fern\'andez et al. (\cite{Fernandez13})  for JFCs. 
As stated in Sect. 2, the determined albedo of ACOs may actually be smaller as we assumed $G = 0.15$ (see Pravec et al. \cite{Pravec2012})  because $H$ may be underestimated for small objects ($H > 12$). In the extreme case that all the $H$ values that we used were 0.5 magnitudes too small (the maximum average difference found by Pravec et al.), all our $p_V$ would be 63\% smaller, and our average $\bar{p_V} = 0.049 $ would drop down to 0.03. In any case, this bias would also affect the comet nuclei albedo determinations based on $H$, so this would not change the main conclusion that both populations have similar albedo distributions.

The $\eta$ distribution of Damocloids looks slightly different than that of JF-ACOs, with a mean beaming parameter $\bar{\eta} =0.94 \pm 0.14$ for the Damocloids and $\bar{\eta} =1.05 \pm 0.16$ for the JF-ACOs. This difference is related to the higher phase angles at which the JF-ACOs were observed by WISE, however, and disappear when we apply the linear relation with $\theta$ to the computed $\eta$. For the entire sample of ACOs, the mean beaming parameter is $\bar{\eta} =1.02 \pm 0.16$, which is compatible with the $\bar{\eta} =1.0$ of the main belt asteroids (Masiero et al. \cite{Masiero2011}) and average beaming parameter of the JFC population $\bar{\eta} =1.03 \pm 0.11$ reported by Fern\'andez et al. (\cite{Fernandez13}).

We conclude that the ACOs in Tancredi (\cite{Tancredi2014}) are likely dormant comets considering dynamical and surface properties criteria, and that Tancredi's dynamical criteria to select dormant comet candidates are much better than those used before. 

We also compare  the cumulative size distribution (CSD) of JF-ACOs, Damocloids, Centaurs, and JFCs. The linear part in the log-log plot of the CSDs for JF-ACOs and Centaurs are almost parallel with a similar cumulative exponent ($\beta = 1.85 \pm 0.30 , = 1.76 \pm 0.35$, respectively), with the CSD of Centaurs shifted towards larger objects than that of the JF-ACOs. The CSD for Damocloids presents a much shallow exponent $\beta = 0.89 \pm 0.17$. The similarity in the exponent between JF-ACOs and Centaurs and its difference with the Damocloids's reinforce the hypothesis that the populations of the first two has a common origin in the transneptunian belt, while the Damocloids comes from the Oort Cloud. 

Finally, the CSD for JF-ACOs is shallower and shifted towards larger diameters with respect to the CSD of active Jupiter family comets; there is only one JFC nuclei larger than 20 km, but there are many JF-ACOs larger than this size. The shallower slope for the JF-ACOs' CSD means that, in relative numbers, there are fewer small members in the JF-ACO population than in  JFC population. In the framework of the hypothesis that the JF-ACOs are dormant comets coming from the JFCs population, this result suggests that the mantling process has a size dependency.  Two effects might be in action. First, JFCs might die by catastrophic disintegration, and this process should be more relevant for the smaller objects.  Second, large comets tend to produce a thick crust, quenching the activity faster than small objects, due to the stronger gravitation field acting on the ejected particles, as suggested by Rickman et al. (\cite{Rickman1990}), Licandro et al. (\cite{Licandro2000}), and Tancredi et al. (\cite{Tancredi2006}). 

We also show that the population of JF-ACOs is comparable in number with the population of JFCs, although there are more tens-km JF-ACOs than JFCs.

%In this paper we present 

%\begin{table}
%\centering
%\begin{tabular}{l c c c c c}
%\hline
%\hline
%Date (UT) & $r$ (AU) & $\Delta$ (AU)  & $\alpha$($^{\circ}$) & $Sun_{PA}$($^{\circ}$) & $Motion_{PA}$($^{\circ}$) \\ \hline
%29/11/2011 & 2.52 & 1.67 & 14.0 & 65.2 & 247.5 \\  \hline
%\end{tabular}
%  \caption{Observing geometry. $r$ and $\Delta$ are the heliocentric and geocentric distance respectively, $\alpha$ is the phase angle, $Sun_{PA}$ and $Motion_{PA}$ are the position angles of the extended Sun--target radius vector and the negative of the target's heliocentric velocity vector}
%  \label{Table1}
%\end{table}

%%%%%%%%%%%%%%%%%%%%%%%%%%%
%%%%%%%%%%%%%%%%%%%%%%%%%%%

\begin{acknowledgements}
We acknowledge James Bauer for his useful comments, which  improved this manuscript. VA-L and JL acknowledge support from the project AYA2012-39115-C03-03 and ESP2013-47816-C4-2-P (MINECO, Spanish Ministry of Economy and Competitiveness). 
  This publication uses data products from NEOWISE, a project of the Jet Propulsion Laboratory/California Institute of Technology, funded by the Planetary Science Division of the NASA. 
  We made use of the NASA/ IPAC Infrared Science Archive, which is operated by the Jet Propulsion Laboratory, California Institute of Technology, under contract with the NASA.  
\end{acknowledgements}

%\bibliographystyle{aa}
%\bibliography{AsteroidsGeneral}

\longtab{1}{
  \begin{longtable}{l c c  c  c  c  c  c c c c}
    \caption{ \label{tabJF1} Observational circumstances of the ACOs in JFC orbits. We also include each object's absolute magnitude $H$, assumed slope parameter $G$, and Tisserand's parameter $T_J$. Columns ``W$n$" ($n=$1 to 4) list the number of useful images in that particular WISE band. Final columns show the heliocentric distance $r$ in AU, distance of the asteroid to WISE spacecraft $\Delta$ in AU, and phase angle $\theta$ in degrees at the midpoint time of the observation.}\\
    \hline
    Object                              &$H$   &$G$     &$T_J$  &W1 &W2 &W3 &W4 &$r$  &$\Delta$ &$\theta$\\
    \hline
    \endfirsthead
    \caption{continued.}\\
    \hline
    Object                              &$H$   &$G$     &$T_J$  &W1 &W2 &W3 &W4 &$r$  &$\Delta$ &$\theta$\\
    \hline
    \endhead
    \hline
    \endfoot    
    (944) Hidalgo                       &10.77 &0.15    &2.067  &0 &0 &5 &5 &9.067  &9.014   &6.262\\
    (5164) Mullo                                &13.00 &0.15    &2.787 &0 &0 &5 &4 &5.347   &5.248  &10.904\\
    (6144) Kondojiro                    &11.50 &0.15    &2.867  &0 &0 &10 &10 &6.489   &6.354   &8.705 \\
    (20898) Fountainhills               &11.10 &0.15    &2.349  &11 &11 &0 &15 &2.270   &1.974  &26.234 \\ 
    (30512) 2001 HO$_8 $                &12.50 &0.15    &2.819  &0 &0 &10 &10 &4.484  &4.357  &13.021\\
    (32511) 2001 NX$_{17}$      &12.80 &0.15    &2.791 &0 &0 &10 &10 &5.324   &5.144  &10.878\\
    (52007) 2002 EQ$_{47}$      &13.30 &0.15    &2.923 &0 &0 &11 &10 &4.354   &4.155  &13.342\\
    (70032) 1999 CZ$_{13}$      &13.20 &0.15    &3.011  &1 &0 &18 &15 &4.757   &4.650  &12.335\\
    (90572) 2004 GN$_{17}$      &13.80 &0.15    &2.991 &0 &0 &11 &11 &3.869   &3.635  &15.075\\
    (96177) 1984 BC                     &16.00 &0.15    &2.780  &23 &13 &24 &23 &1.644   &1.219  &37.721\\
    (119039) 2001 FZ$_{91}$     &13.30 &0.15    &2.933 &0 &0 &4 &4 &3.548   &3.296  &16.461\\
    (145627) 2006 RY$_{102}$    &11.00 &0.15    &2.830  &2 &0 &13 &13 &4.943   &4.750  &11.647\\
    (228862) 2003 GX$_{31}$     &14.30 &0.15    &2.937  &0 &0 &6 &3 &4.736   &4.546  &11.978\\
    (237838) 2002 EV$_{71}$     &14.80 &0.15    &2.906  &0 &0 &14 &12 &2.938   &2.691  &19.807\\
    (248590) 2006 CS            &16.60 &0.15    &2.441  &0 &0 &4 &1 &2.974   &2.811  &19.963\\
    (249047) 2007 TC$_{91}$     &14.10 &0.15    &2.944  &0 &0 &9 &5 &4.249   &4.040  &13.590\\
    (275618) 2000 AU$_{242}$    &13.90 &0.15    &2.736 &0 &0 &11 &10 &3.293   &3.046  &17.493\\
    (301964) 2000 EJ$_{37}$     &13.50 &0.15    &2.441  &0 &1 &9 &10 &3.204   &2.957  &18.383\\
    (306173) 2010 NK$_{83}$     &13.70 &0.15    &2.569  &0 &0 &11 &11 &3.795   &3.674  &15.532\\
    (306461) 1999 LY$_5$        &14.80 &0.15    &2.887 &0 &0 &14 &13 &2.889  &2.719  &20.415\\
    2003 BM$_1$                         &18.30 &0.15    &2.840  &0 &0 &14 &0 &1.877   &1.502  &32.355\\
    2003 BU$_{35}$                      &16.40 &0.15    &2.773 &8 &18 &33 &26 &1.759   &1.370  &34.050\\
    2004 EB                             &17.20 &0.15    &2.755    &0   &0  &13   &3 &2.132  &1.885  &28.505\\ 
    2005 EZ$_{92}$                      &17.30 &0.15    &3.021  &0 &13 &19 &15 &1.506   &1.061  &40.709\\
    2005 NA$_{56}$                      &16.90 &0.15    &2.983  &0 &0 &15 &0 &1.565   &1.214  &39.692\\
    2008 SK$_{220}$                     &15.30 &0.15    &3.005  &0 &0 &6 &2 &3.457   &3.233  &16.746\\
    2008 UD$_{253}$                     &14.30 &0.15    &2.721  &0 &0 &9 &5 &4.068   &3.848  &14.213\\
    2009 DK$_{12}$                      &18.30 &0.15    &2.352 &0 &0 &7 &3 &2.994   &2.816  &19.838\\
    2009 SR$_{143}$                     &16.40 &0.15    &2.806 &0 &0 &4 &0 &2.814   &2.586  &20.436\\
    2009 UA$_{17}$                      &15.90 &0.15    &2.801 &0 &0 &21 &18 &2.485   &2.206  &23.258\\
    2009 WF$_{104}$             &17.20 &0.15    &2.799  &0 &9 &13 &11 &1.807   &1.434  &32.928\\
    2009 WO$_6$                         &17.30 &0.15   &2.784    &17  &11  &18  &18  &1.363  &0.876  &45.908\\ 
    2009 XY$_{21}$                      &15.60 &0.15    &2.815 &0 &0 &11 &0 &3.215   &2.963  &18.083\\
    2010 AN$_{39}$                      &15.50 &0.15    &2.861  &0 &0 &9 &6 &3.127   &2.887  &18.563\\
    2010 AZ$_{60}$                      &15.70 &0.15    &2.737 &0 &0 &13 &11 &2.869   &2.605  &20.335\\
    2010 BD$_{15}$                      &15.70 &0.15    &2.752  &0 &0 &8 &6 &3.020   &2.785  &18.965\\
    2010 CU$_{139}$             &17.30 &0.15    &2.818  &0 &0 &14 &4 &2.377   &2.057  &25.171\\
    2010 HQ                             &15.70 &0.15    &2.750 &0 &0 &2 &1 &4.706   &4.590  &12.311\\
    2010 KG$_{43}$                      &14.50 &0.15    &2.697   &0   &0   &7   &1 &2.934  &2.752  &20.170\\   
    2010 LR$_{68}$                      &18.30 &0.15    &2.923 &0 &0 &18 &3 &1.943   &1.641  &31.491\\
    2010 LV$_{121}$                     &14.80 &0.15   &2.694  &0  &0 &16 &16 &2.974  &2.789  &19.955\\  
    2010 MK$_{43}$                      &15.60 &0.15    &2.888 &0 &0 &15 &14 &2.361   &2.137  &25.493\\
    2010 CC$_{132}$                     &15.90 &0.15    &2.851 &0 &17 &21 &19 &1.894   &1.609  &31.425\\
    2010 DD$_2$                         &17.30 &0.15    &2.930 &0 &0 &4 &0 &1.674   &1.254  &36.883\\
    2010 DD$_{58}$                      &16.30 &0.15    &2.619 &0 &2 &11 &8 &2.033   &1.709  &29.063\\
    2010 FJ$_{81}$                      &20.80 &0.15    &2.340  &0 &13 &46 &2 &1.240   &0.703  &55.814\\
    2010 LB$_{71}$                      &16.20 &0.15    &2.809 &0 &0 &9 &0 &2.984 &2.791  &19.883\\
    2010 MB$_{52}$                      &17.90 &0.15    &2.805  &0 &1 &8 &2 &1.876   &1.583  &32.806\\
    2010 OE$_{101}$                     &17.80 &0.15    &2.868 &50 &13 &65 &64 &1.377   &0.940  &47.479\\
  \end{longtable}
}

\longtab{3}{
  \begin{longtable}{l c c c c}
    \caption{\label{tabJF2} Results of thermal fits and albedo calculations for the ACOs in JFC orbits. We list the best-fitting diameter $D$, the calculated geometric albedo $p_V$, the best-fitting beaming parameter $\eta$ (where applicable), and the near-infrared albedo ratio $p_{\mathrm{W1}}$ (where applicable).}\\
    \hline
    Object                                 &$D (km)$   &$p_V$      &$\eta$ &$p_{\mathrm{W1}}$ \\
    \hline
    \endfirsthead
    \caption{continued.}\\
    \hline
    Object                                 &$D (km)$   &$p_V$      &$\eta$ &$p_{\mathrm{W1}}$ \\
    \hline
    \endhead
    \hline
    \endfoot     
    (944) Hidalgo                       &61.4 $\pm$ 12.7        &0.028 $\pm$ 0.006           &0.90 $\pm$ 0.24 &--- \\ 
    (5164) Mullo                                &15.4 $\pm$ 2.0         &0.050 $\pm$ 0.010             &0.96 $\pm$ 0.21 &---    \\ 
    (6144) Kondojiro                    &32.9 $\pm$ 5.1         &0.044 $\pm$ 0.009           &0.93 $\pm$ 0.20 &---  \\ 
    (20898) Fountainhills               &41.53 $\pm$ 0.85       &0.037 $\pm$ 0.007           &---                            &0.093 $\pm$ 0.010   \\ 
    (30512) 2001 HO$_8 $                &18.3 $\pm$ 2.9         &0.057 $\pm$ 0.011           &1.01 $\pm$ 0.16 &---  \\ 
    (32511) 2001 NX$_{17}$      &16.3 $\pm$ 2.7         &0.055 $\pm$ 0.011                 &1.02 $\pm$ 0.26 &---   \\ 
    (52007) 2002 EQ$_{47}$      &11.5 $\pm$ 1.7         &0.067 $\pm$ 0.013                 &1.03 $\pm$ 0.26 &---  \\ 
    (70032) 1999 CZ$_{13}$      &14.1 $\pm$ 3.0         &0.052 $\pm$ 0.010                 &1.00 $\pm$ 0.34 &---  \\ 
    (90572) 2004 GN$_{17}$      &12.1 $\pm$ 2.0         &0.038 $\pm$ 0.008                 &1.15 $\pm$ 0.22 &---  \\ 
    (96177) 1984 BC                     &3.49 $\pm$ 0.24        &0.059 $\pm$ 0.012           &1.05 $\pm$ 0.13 &0.132 $\pm$ 0.081      \\ 
    (119039) 2001 FZ$_{91}$     &14.08 $\pm$ 0.43       &0.043 $\pm$ 0.009                 &0.87 $\pm$ 0.03 &---  \\ 
    (145627) 2006 RY$_{102}$    &37.0 $\pm$ 1.0         &0.051 $\pm$ 0.010                 &0.83 $\pm$ 0.04 &---  \\ 
    (228862) 2003 GX$_{31}$     &10.58 $\pm$ 0.66       &0.030 $\pm$ 0.006                 &--- &---  \\ 
    (237838) 2002 EV$_{71}$     &6.51 $\pm$ 0.87        &0.053 $\pm$ 0.011                 &1.09 $\pm$ 0.18 &---   \\ 
    (248590) 2006 CS            &3.29 $\pm$ 0.03        &0.038 $\pm$ 0.008                 &--- &---  \\ 
    (249047) 2007 TC$_{91}$     &9.7 $\pm$ 1.9  &0.048 $\pm$ 0.010              &1.10 $\pm$ 0.30 &---   \\ 
    (275618) 2000 AU$_{242}$    &8.59 $\pm$ 0.90        &0.068 $\pm$ 0.014                 &0.99 $\pm$ 0.21 &---   \\ 
    (301964) 2000 EJ$_{37}$     &12.43 $\pm$ 0.72       &0.046 $\pm$ 0.009                 &0.95 $\pm$ 0.07 &---  \\ 
    (306173) 2010 NK$_{83}$     &11.6 $\pm$ 2.1         &0.048 $\pm$ 0.010                 &1.18 $\pm$ 0.31 &---  \\ 
    (306461) 1999 LY$_5$        &6.45 $\pm$ 0.97        &0.053 $\pm$ 0.011         &1.00 $\pm$ 0.15 &--- \\ 
    2003 BM$_1$                         &1.21 $\pm$ 0.14        &0.060 $\pm$ 0.012           &--- &---    \\
    2003 BU$_{35}$                      &3.32 $\pm$ 0.31        &0.045 $\pm$ 0.009           &0.97 $\pm$ 0.14 &---    \\ 
    2004 EB                             &2.03 $\pm$ 0.30        &0.058 $\pm$ 0.012           &--- &--- \\ 
    2005 EZ$_{92}$                      &1.97 $\pm$ 0.20        &0.057 $\pm$ 0.011           &1.27 $\pm$ 0.24 &---   \\ 
    2005 NA$_{56}$                      &1.07 $\pm$ 0.05        &0.19 $\pm$ 0.04            &--- &---  \\ 
    2008 SK$_{220}$                     &4.91 $\pm$ 0.32        &0.056 $\pm$ 0.011           &--- &---  \\ 
    2008 UD$_{253}$                     &12.09 $\pm$ 3.05       &0.028 $\pm$ 0.006           &1.63 $\pm$ 0.69 &---  \\ 
    2009 DK$_{12}$                      &5.35 $\pm$ 0.29        &0.047 $\pm$ 0.009           &1.10 $\pm$ 0.04 &---   \\ 
    2009 SR$_{143}$                     &3.41 $\pm$ 0.36        &0.043 $\pm$ 0.009           &--- &---  \\ 
    2009 UA$_{17}$                      &4.03 $\pm$ 0.87        &0.053 $\pm$ 0.011           &0.98 $\pm$ 0.37 &---   \\ 
    2009 WF$_{104}$             &2.47 $\pm$ 0.53        &0.046 $\pm$ 0.009                 &1.15 $\pm$ 0.42 &---   \\ 
    2009 WO$_6$                         &2.53 $\pm$ 0.38        &0.036 $\pm$ 0.007   &1.16 $\pm$ 0.17    &0.02 $\pm$  0.19\\
    2009 XY$_{21}$                      &4.22 $\pm$ 0.33        &0.058 $\pm$ 0.012           &--- &---   \\
    2010 AN$_{39}$                      &5.45 $\pm$ 0.85        &0.040 $\pm$ 0.008           &1.03 $\pm$ 0.23 &---    \\ 
    2010 AZ$_{60}$                      &4.40 $\pm$ 0.59        &0.050 $\pm$ 0.010           &0.88 $\pm$ 0.20 &---    \\ 
    2010 BD$_{15}$                      &5.90 $\pm$ 0.70        &0.028 $\pm$ 0.006           &0.97 $\pm$ 0.18 &---   \\ 
    2010 CU$_{139}$             &2.34 $\pm$ 0.10        &0.039 $\pm$ 0.008                 &--- &---  \\ 
    2010 HQ                             &12.23 $\pm$ 0.67       &0.036 $\pm$ 0.008           &--- &--- \\ 
    2010 KG$_{43}$                      &3.49 $\pm$ 0.52        &0.23 $\pm$ 0.05            &--- &--- \\ 
    2010 LR$_{68}$                      &1.66 $\pm$ 0.07        &0.031 $\pm$ 0.006           &--- &---   \\
    2010 LV$_{121}$                     &5.31 $\pm$ 0.80        &0.081 $\pm$ 0.016   &0.92 $\pm$ 0.14   &---\\
    2010 MK$_{43}$                      &3.73 $\pm$ 0.49        &0.077 $\pm$ 0.015           &1.18 $\pm$ 0.25 &---   \\ 
    2010 CC$_{132}$                     &3.66 $\pm$ 0.37        &0.059 $\pm$ 0.012           &0.94 $\pm$ 0.17 &---   \\ 
    2010 DD$_2$                         &0.79 $\pm$ 0.12        &0.339 $\pm$ 0.068           & --- &---  \\
    2010 DD$_{58}$                      &3.25 $\pm$ 0.56        &0.055 $\pm$ 0.011           &1.19 $\pm$ 0.34 &---    \\ 
    2010 FJ$_{81}$                      &0.44 $\pm$ 0.07        &0.048 $\pm$ 0.010           &--- &---  \\ 
    2010 LB$_{71}$                      &3.44 $\pm$ 0.44        &0.052 $\pm$ 0.010           &--- &---  \\ 
    2010 MB$_{52}$                      &1.70 $\pm$ 0.21        &0.045 $\pm$ 0.009           &--- &---   \\ 
    2010 OE$_{101}$                     &2.14 $\pm$ 0.18        &0.030 $\pm$ 0.006           &1.30 $\pm$ 0.22 &0.17 $\pm$ 0.13        \\ 
    \hline
  \end{longtable}
}

\end{document}